\documentclass[twocolumn,natbib]{svjour3}   
\pdfoutput=1
\smartqed  
\usepackage{amsmath}
\usepackage{amssymb}
\usepackage{mathrsfs}
\usepackage{graphicx}
\usepackage{subfigure}
\usepackage{epstopdf}
\usepackage{placeins}
\usepackage[activate={true,nocompatibility},final,tracking=true,kerning=true,spacing=true,factor=1100,stretch=10,shrink=10]{microtype}
\usepackage{xcolor}
\usepackage[expproduct=cdot]{siunitx}

\usepackage{booktabs}

\usepackage{bbm} 
\usepackage{upgreek}\newcommand{\betaup}{\upbeta}
\usepackage{relsize} 

\renewcommand{\b}[1]{{\boldsymbol{#1}}} 
\newcommand{\der}{\mathrm{d}}

\newcommand{\tpd}[2]{\tfrac{\partial #1}{\partial #2}}

\newcommand{\fbm}{\ensuremath{{f_\text{bm}}}}
\newcommand{\fvas}{\ensuremath{{f_\text{vas}}}}
\newcommand{\bm}{\text{bm}}
\newcommand{\vas}{\text{vas}}
\newcommand{\rve}{\text{RVE}}
\newcommand{\bv}{\text{BV}}
\newcommand{\tv}{\text{TV}}
\newcommand{\Ctissue}{\ensuremath{\mathbb{C}^\text{tissue}}}
\newcommand{\tissue}{\text{tissue}}
\newcommand{\micro}{\text{micro}}
\newcommand{\cmicro}{\ensuremath{\mathbbm{c}^\text{micro}}}
\newcommand{\sedtissue}{\ensuremath{\Psi^\tissue}}
\newcommand{\sedmicro}{\ensuremath{\Psi_\bm^\micro}}
\newcommand{\sv}{\ensuremath{\text{S}_{\textrm{V}}}}
\newcommand{\da}{\ensuremath{\text{day}}}

\newcommand{\ob}{\text{\textsmaller{OB}}}

\newcommand{\obu}{\text{\textsmaller{OB}$_\text{u}$}} 

\newcommand{\obp}{\text{\textsmaller{OB}$_\text{p}$}}

\newcommand{\oba}{\text{\textsmaller{OB}$_\text{a}$}}

\newcommand{\ocu}{\text{\textsmaller{OC}$_\text{u}$}}
\newcommand{\ocp}{\text{\textsmaller{OC}$_\text{p}$}}

\newcommand{\oca}{\text{\textsmaller{OC}$_\text{a}$}}

\newcommand{\tgfb}{\text{\textsmaller{TGF\textsmaller{$\betaup$}}}}

\newcommand{\wnt}{\text{\textsmaller{Wnt}}}

\newcommand{\mcsf}{\text{\textsmaller{MCSF}}}
\newcommand{\nfkb}{\text{\textsmaller{NF$\kappa$B}}}
\newcommand{\rank}{\text{\textsmaller{RANK}}}
\newcommand{\rankl}{\text{\textsmaller{RANKL}}}

\newcommand{\opg}{\text{\textsmaller{OPG}}}

\newcommand{\pth}{\text{\textsmaller{PTH}}}

\newcommand{\pirep}{\ensuremath{\pi^\text{rep}}} 
\newcommand{\piact}{\ensuremath{\pi^\text{act}}} 
\newcommand{\kform}{\text{$k_\text{form}$}} 

\newcommand{\kres}{\text{$k_\text{res}$}}

\newcommand{\dobu}{\ensuremath{\mathcal{D}_\obu}}
\newcommand{\dobp}{\ensuremath{\mathcal{D}_\obp}}

\newcommand{\pobp}{\ensuremath{\mathcal{P}_\obp}}
\newcommand{\docu}{\ensuremath{\mathcal{D}_\ocu}}
\newcommand{\docp}{\ensuremath{\mathcal{D}_\ocp}}
\newcommand{\aoca}{\ensuremath{\mathcal{A}_\oca}}

\makeatletter
\newsavebox{\@tabnotebox}
\providecommand\tmark{} 
\providecommand\tnote{}
\newenvironment{tabularwithnotes}[3][c]
  {\long\def\@tabnotes{#3}%
   \renewcommand\tmark[1][a]{\makebox[0pt][l]{\textsuperscript{##1}}}%
   \renewcommand\tnote[2][a]{\textsuperscript{##1}\,##2\par}
   \begin{lrbox}{\@tabnotebox}
   \begin{tabular}{#2}}
  {\end{tabular}\end{lrbox}%
   \parbox{\wd\@tabnotebox}{
     \usebox{\@tabnotebox}\par
     \smallskip\@tabnotes
   }%
  }
\makeatother

\begin{document}

\title{A multiscale mechanobiological model of bone remodelling predicts site-specific bone loss in the femur during osteoporosis and mechanical disuse}

\author{C. Lerebours	\and P. R. Buenzli \and S. Scheiner \and  P. Pivonka}

\titlerunning{A multiscale mechanobiological model of bone remodelling}        


\institute{C. Lerebours - P. R. Buenzli \at
              School of Mathematical Sciences, Monash University,\\
              Clayton, VIC, Australia. \\
              Tel.: +613-99-024002\\
              \email{chloe.lerebours@monash.edu} 
           \and
           S. Scheiner \at
           	Institute for Mechanics of Materials and Structures, Vienna University of Technology, Vienna, Austria.          
           \and
           P. Pivonka \at
              Northwest Academic Centre, University of Melbourne,\\
              St Albans, VIC, Australia.  
}

\date{Received: date / Accepted: date}

\maketitle

\begin{abstract}
We propose a multiscale me\-chano\-bio\-logical model of bone remodelling to investigate the site-specific evolution of bone volume fraction across the midshaft of a femur. The model includes hormonal regulation and biochemical coupling of bone cell populations, the influence of the microstructure on bone turn\-over rate, and mechanical adaptation of the tissue. Both microscopic and tissue-scale stress/strain states of the tissue are calculated from macroscopic loads by a combination of beam theory and micromechanical homogenisation.

This model is applied to simulate the spatio-temporal evolution of a human midshaft femur scan subjected to two deregulating circumstances: (i) osteoporosis and (ii) mechanical disuse. Both simulated deregulations led to endocortical bone loss, cortical wall thinning and expansion of the medullary cavity, in accordance with experimental findings. Our model suggests that these observations are attributable to a large extent to the influence of the microstructure on bone turnover rate. Mechanical adaptation is found to help preserve intracortical bone matrix near the periosteum. Moreover, it leads to non-uniform cortical wall thickness due to the asymmetry of macroscopic loads introduced by the bending moment. The effect of mechanical adaptation near the endosteum can be greatly affected by whether the mechanical stimulus includes stress concentration effects or not.

\end{abstract}

\section{Introduction}
\label{intro}
\begin{figure*}[t!]
\begin{center}
\includegraphics[trim = 0cm 7.5cm 0.3cm 7cm, clip=true,  width = \textwidth]{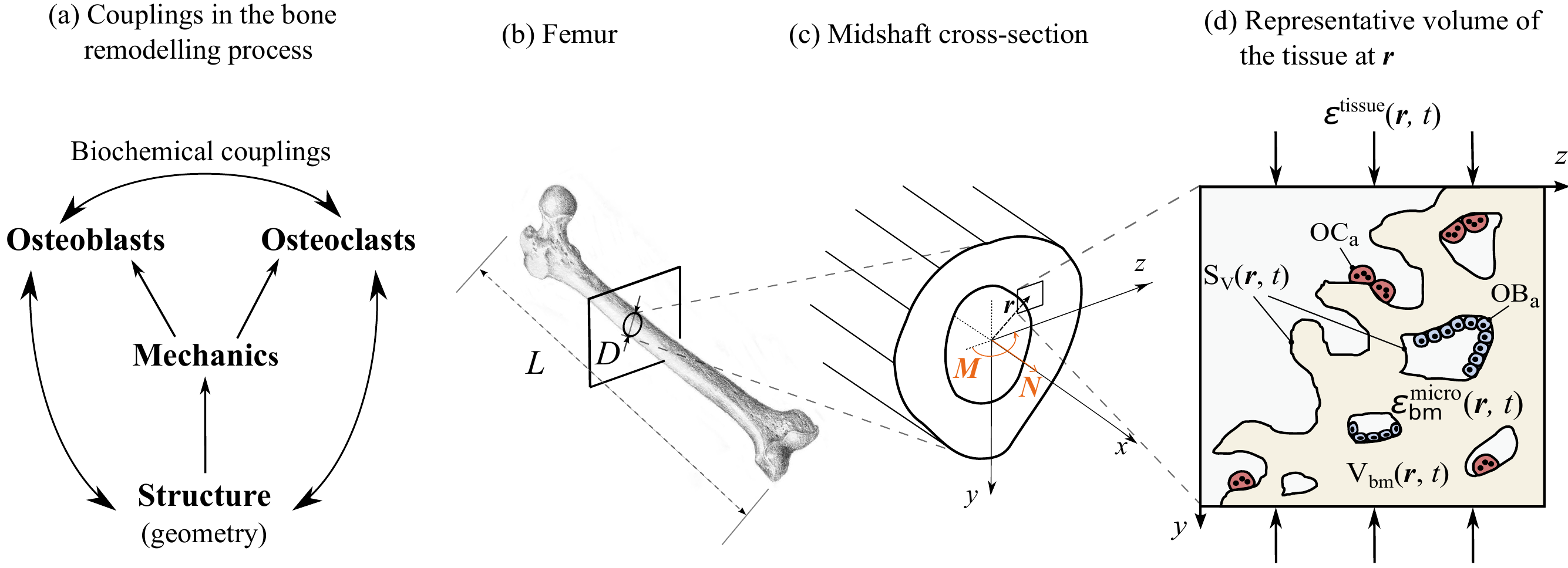} 
\caption{Multiscale representation of bone. (a) Scheme of the couplings in the bone remodelling process; (b) Femur bone geometry (organ scale); (c) Midshaft cross section depicting coordinate axes and the sectional forces used for beam theory (tissue to organ scales); (d) Representative volume element (RVE) of cortical bone used to define bone cell densities, bone volume fraction, and specific surface (cellular to tissue scales).}
\label{multiscale}
\end{center}
\end{figure*}

Bone is a biomaterial with a complex hierarchical structure characterised by at least three distinct length scales: (i) the cellular scale (10--20 $\mu$m); (ii) the tissue scale (2--5 mm) and (iii) the whole organ scale (4--45 cm) [\cite{Rho1998a, Weiner1998}]. Several interactions exist between these scales, which affect bone remodelling, bone material properties and bone structural integrity. The activity of bone-resorbing and bone-forming cells during bone remodelling leads to changes in material properties at the tissue scale which subsequently affect the distribution of loads at the structural, whole organ scale (Figure \ref{multiscale}). Besides, changes in bone shape and microarchitecture modify the stress/strain distribution and bone surface availability, which provide mechanical and geometrical feedbacks onto the bone cells and, eventually, affect bone remodelling [\cite{Martin1972, Lanyon1982, Frost1987}]. Due to the complexity of these interactions, the interpretation of experimental data at a single scale is difficult. Predicting the evolution of multifactorial bone disorders, such as osteoporosis, necessitates a comprehensive modelling approach in which these multiscale interactions are consistently integrated. 

Various mathematical models of bone remodelling have been proposed in the literature. Biomechanics models estimating tissue-scale stress and strain distribution from musculoskeletal models and average material properties, such as bone density, are often used in conjunction with remodelling algorithms based on Wolff's law. These remodelling algorithms locally increase or decrease bone density depending on the tissue's mechanical state [\cite{Carter1977, Carter2001, Fyhrie1986, Weinans1992, Meulen1993, Pettermann1997}]. Such models may also include damage accumulation due to fatigue loading and damage repair [\cite{Prendergast1994, McNamara2007, Garcia-Aznar2005}]. Other models focus at the microstructural scale ($\mu$m\ to mm) and describe the evolution of trabecular bone microarchitecture through resorption and formation events at the bone surface induced by the local mechanical state [\cite{Huiskes2000, Ruimerman2005, VanOers2008, Christen2012, Christen2013a}]. Most of these mathematical models focus on the biomechanical aspects of bone remodelling and do not consider hormonal regulation or biochemical coupling between bone cells.

In this paper, we propose a novel multiscale modelling approach of bone remodelling combining and extending several mathematical models into a consistent framework. This framework enables (i) the consideration of biochemical and cellular interactions in bone remodelling at the cellular scale [\cite{Lemaire2004, Pivonka2008, Buenzli2013a, Pivonka2013}], (ii) the evolution of material properties at the tissue scale based on bone cell remodelling activities regulated by mechanical feedback [\cite{Scheiner2013}] and bone surface availability [\cite{Pivonka2013, Buenzli2013}], and (iii) the determination of the stress/ strain distributions from the tissue scale to the microstructural scale by a combination of generalised beam theory and micromechanical homogenisation [\cite{Hellmich2008, Scheiner2013, Buenzli2013}]. 

This modelling approach is applied to simulate the temporal evolution of a human femoral bone at the midshaft (Figure \ref{multiscale}), subjected to various deregulating circumstances such as osteoporosis and changes in mechanical loading. An initial state of normal bone remodelling is first assumed, in which the tissue across the midshaft cross section remodels at site-specific turnover rates without changing its average material properties. Osteoporosis is then simulated by hormonal changes deregulating the biochemical coupling between osteoclasts and osteoblasts. These hormonal changes are calibrated so as to reproduce realistic rates of osteoporotic bone loss. The strength of the resorptive and formative responses of bone cells to mechanical feedbacks are calibrated so as to reproduce rates of bone loss and recovery in cosmonauts undertaking long-duration space flight missions. A scan of a femur cross section is used as initial condition for our simulations. This illustrates the potential of our modelling approach to be used as a predictive, patient-specific diagnostic tool for estimating the deterioration of bone tissues. Here, we use the model to investigate the interplay between geometrical and mechanical feedbacks in inducing site-specific bone loss in osteoporosis, which is characterised by endocortical bone loss, cortical wall thinning, and the expansion of the marrow cavity [\cite{Feik1997, Bousson2001, Zebaze2010}].

\section{Description of the model}\label{Mat_Methods}
\begin{figure*}[t!]
\begin{center}
\includegraphics[trim = 0.4cm 6cm 0.4cm 2.5cm, clip=true,  width = 0.85\textwidth]{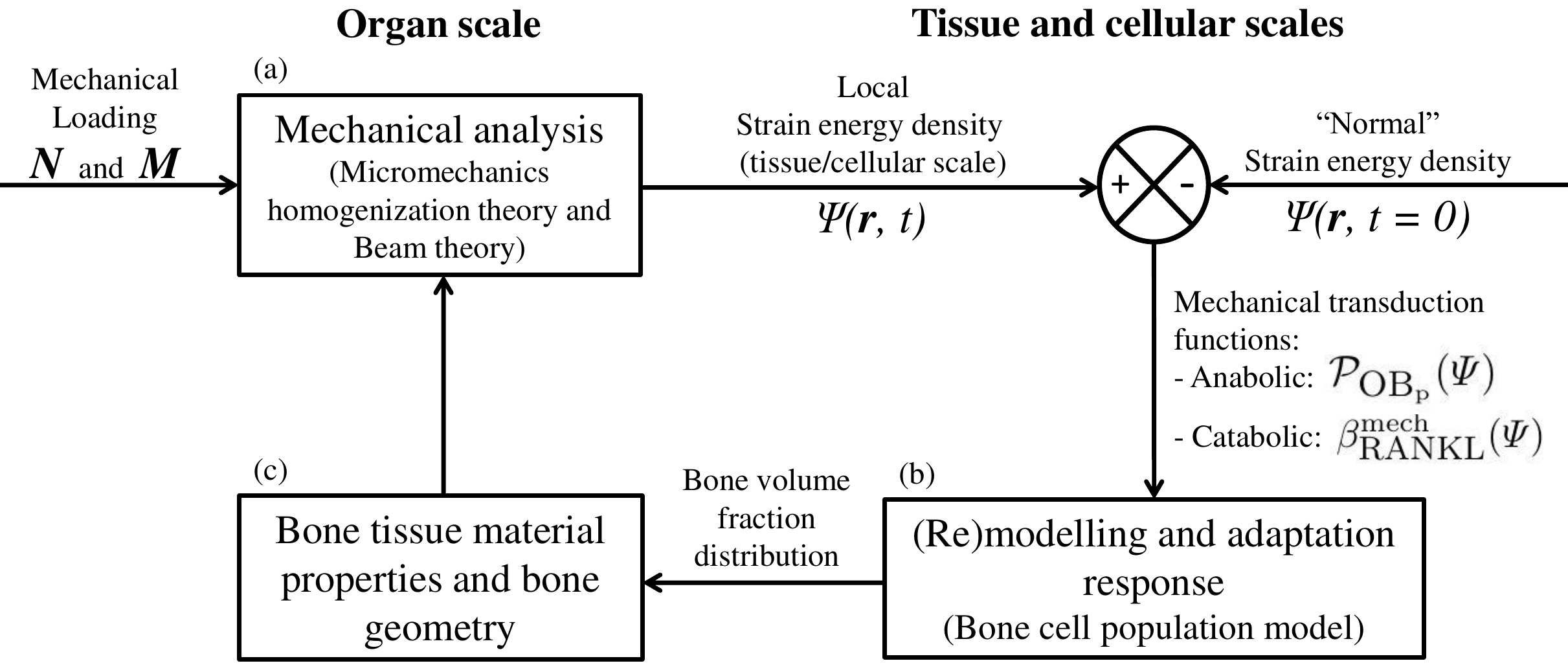} 
\caption{Flow chart of bone remodelling simulations taking into account (a) the global mechanical loading, (b) the bone cell population model, and (c) the bone material and geometry adaptation.}
\label{model}
\end{center}
\end{figure*}

Figures \ref{multiscale} and \ref{model} summarise the general approach of our model. We consider a portion of human femur near the midshaft. This portion of bone is assumed to carry loads corresponding to a total normal force $\b N$ and total bending moment $ \b M$ (Figure~\ref{multiscale}(c)). These loads are distributed unevenly across the midshaft cross section depending on the site-specific bone microstructure, particularly on the cortical porosity [\cite{Zebaze2010, Buenzli2013}]. This load distribution determines a site-specific mechanical stimulus which is sensed and transduced by bone cells (Figure \ref{model}(a)). This mechanical feedback is incorporated in a cell population model as biochemical signals leading to changes in the balance between osteoclasts and osteoblasts (Figure \ref{model}(b)). In addition, microstructural parameters such as bone volume fraction (\fbm) and bone specific surface influence the propensity of bone cells to differentiate and become active [\cite{Martin1984, Lerebours2014}]. This geometrical feedback is included in the cell population model via a dependence of the bone turnover rate on the bone volume fraction. The activities of osteoclasts and osteoblasts modify the tissue microstructural parameters (bone volume fraction, bone specific surface), which in turn induces changes in the load distribution (Figure \ref{model}(c)). In the following, we introduce in more detail the multiple scales and related variables involved in this model workflow. Table \ref{nomenclature}, in Appendix \ref{table_section}, lists all the parameters of the model.

\subsection{Load distribution from the organ scale to the cellular scale}\label{methods_organ_to_cellular}
Loading is composed of body weight and muscle forces exerted onto bone via tendons and direct action of muscles. These forces can be calculated from bone shape, muscle and tendon attachment, and gait analysis data using musculoskeletal models~[\cite{Lloyd2003, Viceconti2006, Martelli2014}]. Continuum mechanics provides the link between external forces exerted onto a structure, and the strain and stress distribution in the structure [\cite{Salencon2001}].

Tissue-scale properties within the framework of continuum mechanics are average mechanical properties over microstructural material phases and pores (presented in detail in the next sections). The corresponding tissue-scale stresses and strains may significantly deviate from the microscopic, cellular scale, stresses and strains acting in the different material phases composing the tissue due to so-called strain and stress concentration effects [\cite{Zaoui2002, Hill1963}]. Microscopic stress and strain distributions in the bone matrix are likely to be sensed directly by bone cells, particularly by osteocytes [\cite{Scheiner2013}]. However, as osteocytes form an extensive interconnected network [\cite{Marotti2000, Buenzli2015}], they may also sense larger scale stress and strain distributions. We will let either the tissue-scale or the microscopic mechanical state of bone act onto the bone cells to investigate how this influences the site-specific evolution of bone tissue microstructures.

In the following, we present first how stress and strain distributions can be calculated at the tissue scale using beam theory. We then present how these tissue-scale stress and strain distributions are employed as site-specific loading boundary conditions to the continuum micromechanical model of \cite{Hellmich2008} for the calculation of microscopic stress and strain distributions effective at the cellular level.

\subsubsection*{Determination of tissue-scale stress and strain distributions based on beam theory}
The continuum mechanical field equations allow the calculation of tissue-scale strain and stress distributions in bone. Given that the length of the femur $L$ is significantly larger (45--50 cm) than its diameter $D$ (3--5 cm) at the midshaft (Figure \ref{multiscale}(b)) the continuum mechanical field equations can be approximated using beam theory formulated for small strains and generalised to materials of non-uniform properties, an approach we have used previously in~\cite{Buenzli2013}. 

This approach requires the knowledge of the total external forces, i.e., the normal force $\b N$ and the bending moment $\b M$ carried by the femur cross section. $\b N$ and $\b M$ can be estimated for different physical activities by using musculoskeletal models [\cite{Vaughan1992, Forner-Cordero2006}]. In our simulations we take constant values for $\b N$ and $\b M$ comparable with the maximum ground reaction force and knee and hip moments that occur during a gait analysis, estimated as: $\b N = (N_{x}, 0, 0), ~N_{x} = - 700 $ N, and $\b M = M ~\hat{\b m}$, M = 50 Nm, where $\hat{\b m}$ is a unit vector along the antero-posterior axis of the cross section determined from the micro-radiograph [\cite{Vaughan1992, Forner-Cordero2006, Ruff2000, Cordey1999}] (Figure \ref{extraction}(c)). The $x$-axis is the femur's longitudinal axis and ($y$,$z$) is the plane transverse to $x$ at the midshaft (Figure~\ref{multiscale}(b)-(c)).

Tissue-scale mechanical properties correspond to spatial averages over a so-called representative volume element (RVE) of the tissue. In cortical bone, an appropriate tissue RVE is of the order of $10\times2\times2 ~\textrm{mm}^{3}$, a size large enough to contain a large number of pores, but small enough to retain site-specific information and to not be influenced by macroscopic features such as overall bone shape~[\cite{Hill1963, Zaoui2002}]. We denote by $\Ctissue(\b r, t)$ the local bone tissue stiffness tensor defined at the RVE scale, where $\b r$ denotes the location in bone of the RVE (Figure~\ref{multiscale}(c)) and the dependence on time $t$ reflects the fact that bone remodelling may modify the local mechanical properties of the tissue. This tissue-scale stiffness tensor is assumed to relate the tissue-scale stress tensor $\b \sigma^\tissue$ and strain tensor $\b \varepsilon^\tissue$ pointwise according to Hooke's law:
\begin{align}\label{hooke}
    \b \sigma^\tissue(\b r, t) = \Ctissue(\b r, t):\b \varepsilon^\tissue(\b r, t).
\end{align}
Beam theory is based on the so-called Euler--Bernoulli kinematic hypothesis, which asserts that material cross sections initially normal to the beam's neutral axis remain planar, undeformed in their own plane, and normal to the neutral axis in the beam's deformed state~[\cite{Timoshenko1951, Bauchau2009, Hjelmstad2005}]. These assumptions are expected to be well satisfied near the femoral midshaft under small deformations generated by bending and compression or tension. Furthermore, no shear force, torsional loads or twisting along the beam axis are assumed. These assumptions, Eq.~\eqref{hooke}, and the fact that bone is an orthotropic material [\cite{Hellmich2004}] imply that the only nonzero components of the stress tensor are the normal stresses $\sigma^\tissue_{xx}=\Ctissue_{1111}\varepsilon^\tissue_{xx}$, $\sigma^\tissue_{yy}=\Ctissue_{1122}\varepsilon^\tissue_{xx}$, and $\sigma^\tissue_{zz}=\Ctissue_{1133}\varepsilon^\tissue_{xx}$, where the normal stresses $\sigma^\tissue_{yy}$ and $\sigma^\tissue_{zz}$ are induced by compression or tension along the beam axis $x$ by the Poisson effect\footnote{The stress components $\sigma^\tissue_{yy}$ and $\sigma^\tissue_{zz}$ do not participate directly to the transfer of the resultant force $\b N$ and resultant bending moment $\b M$ across the bone cross section, however, they are accounted for in the calculation of the tissue-scale strain energy density $\sedtissue$.} (see ~\cite{Buenzli2013} for more details). The Euler--Bernoulli hypothesis implies that the tissue strain tensor reduces to the single non-zero component $\varepsilon^\tissue_{xx}$ and that:
\begin{align}\label{Bernoulli}
    \varepsilon^\tissue_{xx}(y,z,t) = \varepsilon_1(t) - \kappa_3(t) y + \kappa_2(t) z,
\end{align}
where $\varepsilon_1$ is the sectional axial strain, and $\kappa_2$ and $\kappa_3$ are the sectional beam curvatures about the $z$- and $y$-axes, respectively [\cite{Bauchau2009}]. The three unknowns $\varepsilon_1$, $\kappa_2$, and $\kappa_3$ are determined by the constraints that (i) the integral of $\sigma^\tissue_{xx}$ over the midshaft cross section must give the total normal force $N_{x}$, and (ii) the integral of the stress moment $(0,y,z)\times \sigma^\tissue_{xx}\hat{\b x}$ must give the total bending moment $\b M = M ~\hat{\b m}$ (the axes origin in the $(y,z)$ plane is set at the modulus-weighted centroid of the section, also called normal force center~[\cite{Bauchau2009}]). Explicit formulas for $\varepsilon_1, \kappa_2$, and $\kappa_3$ as functions of $\Ctissue$, $\b N$ and $\b M$ are presented in Appendix~\ref{appendix:beam}. We refer the reader to \cite{Bauchau2009}, Sec.~6.3 and \cite{Buenzli2013} for their derivation.

\subsubsection*{Determination of microscopic stress and strain distributions based on micromechanical homogenisation theory}

Bone tissue stiffness $\Ctissue$ is strongly influenced by the tissue's microstructure, in particular its porosity, or equivalently, its bone volume fraction $\fbm$. Bone volume fraction is a microstructural parameter defined at the tissue scale as the volume fraction of bone matrix in the RVE (Figure~\ref{multiscale}(d)):
$
    \fbm = \bv/\tv = 1 - \text{porosity},
$
where $\bv$ is the volume of bone matrix in the RVE and $\tv$ is the tissue volume, i.e. the total volume of the RVE [\cite{Dempster2013}]. In  ~\cite{Buenzli2013}, we used an explicit power-law relationship $\Ctissue_{1111}(\fbm) \propto\fbm^3$ based on experimental relationships between bone stiffness and bone mineral content~[\cite{Carter1977, Hernandez2001}]. While regression approaches based on power-law relations are able to account for material properties in one principal direction, they are less accurate in estimating material properties in other principal directions.   

Here, we follow a different approach taken by Hellmich and colleagues using the framework of continuum micromechanics [\cite{Hill1963, Hill1965, Zaoui1997, Zaoui2002}]. Mechanically, bone tissue can be considered as a two-phase material: a bone matrix phase (`$\bm$') consisting of mineralised bone matrix, and a vascular phase (`$\vas$') consisting of vascular components, cells, extracellular matrix and other soft tissues present in Haversian canals and in the marrow.

Continuum micromechanics provides a framework to estimate the tissue-scale stiffness tensor $\Ctissue(\fbm)$ from the microscopic stiffness properties of bone matrix and vascular pores, and assumptions on pore microarchitecture and phase interactions [\cite{Hellmich2008}]. The advantage of this approach is to provide (i) accurate three-dimensional estimates of $\Ctissue$ and (ii) estimates of the microscopic stress and strain distributions of the bone matrix without recourse to costly micro-finite element analyses of the tissue microstructure~[\cite{Fritsch2009}]. Using the concept of continuum micromechanics is justified in bone due to the separation of length scales between the \rve ~size and the characteristic sizes of the two-phase microstructures  [\cite{Hellmich2008, Scheiner2013}]. We summarise below the premises upon which this approach is based.

The tissue-scale stress and strain tensors $\sigma^\tissue$ and $\varepsilon^\tissue$ correspond to spatial averages over the \rve ~of the microscopic (cellular-scale) stress and strain tensors. Assuming that each phase within the \rve ~is homogeneous, these spatial averages can be expressed as sums over the phases:
\begin{align}
\b \sigma^\tissue(\b r, t) & \equiv \frac{1}{\tv} \int_{\tv} \b \sigma^\micro \der V  =\sum\limits_{k}f_k\, \b \sigma^\micro_k \label{sigma-tissue},
\\\b \varepsilon^\tissue(\b r, t) & \equiv \frac{1}{\tv} \int_{\tv} \b \varepsilon^\micro \der V =\sum\limits_{k}f_k\, \b \varepsilon^\micro_k,
\end{align}
where $f_k(\b r, t)$ is the volume fraction of phase $k$ (`\bm', `\vas'), $\b \sigma^\micro_k(\b r, t)$ and $\b \varepsilon^\micro_k(\b r, t)$ are the microscopic stress and strain tensors in phase $k$. We emphasise that all these quantities still depend on the tissue-scale location $\b r$ of the \rve ~in bone, whilst microscopic inhomogeneities are encoded in the phase index $k$. It can be shown that due to the linearity of the constitutive equations the phase strain tensor  $\b \varepsilon^\micro_k$ is related linearly with the tissue-scale strain tensor: 
\begin{align}
    \b \varepsilon^\micro_k = \mathbb{A}_k : \b \varepsilon^\tissue, \label{micro-macro-strain1}
\end{align} 
where $\mathbb{A}_k$ is a fourth-order tensor called the strain concentration tensor~[\cite{Zaoui2002, Hellmich2008, Fritsch2009}]. Assuming that Hooke's law also holds for each phase at the microscopic scale, $\b \sigma^\micro_k = \cmicro_k : \b \varepsilon^\micro_k$ (with $\cmicro_k$ the stiffness tensor of phase $k$), one obtains from Eqs~\eqref{sigma-tissue} and~\eqref{micro-macro-strain1}:
\begin{align}\label{macro-stiffness-from-micro}
\b \sigma^\tissue & = \sum_k f_k ~\cmicro_k : \b \varepsilon^\micro_k \\
& = \Big(\sum_k f_k ~\cmicro_k: \mathbb{A}_k\Big) : \b \varepsilon^\tissue \equiv \Ctissue : \b \varepsilon^\tissue, \nonumber 
\end{align}
where
\begin{align}
\Ctissue & = \fbm ~\cmicro_\bm:\mathbb{A}_\bm + \fvas ~\cmicro_\vas: \mathbb{A}_\vas. \label{cmacro}
\end{align}
Equation~\eqref{cmacro} provides a relationship between the tissue-scale stiffness, $\Ctissue$, and the microscopic properties of the phases composing the tissue, $f_k, \cmicro_k$, and $\mathbb{A}_k$. Because mineral content across bone tissues only varies little on average~[\cite{Scheiner2013, Fritsch2007}], $\cmicro_\bm$ can be assumed constant and homogeneous, i.e., independent of $\b r, t$. The elastic modulus $\cmicro_\vas$ is likewise assumed independent of $\b r, t$ and taken as that of water~[\cite{Scheiner2013}]. Both $\cmicro_\bm$ and $\cmicro_\vas$ have been measured experimentally, their values are listed in Table \ref{nomenclature}. The strain concentration tensors $\mathbb{A}_k$ can be estimated by solving so-called matrix-inclusion problems of elasticity homogenisation theory, which use assumptions on phase shape within the RVE and phase interactions~[\cite{Eshelby1957, Laws1977}]. For bone, accurate multi-scale homogenisation schemes were developed and validated experimentally [\cite{Hellmich2008, Fritsch2009, Morin2014}]. These schemes provide explicit expressions for $\mathbb{A}_k$ depending on the phase volume fractions $f_\bm$ and $f_\vas$. Because $f_\vas = 1-\fbm$, this defines both the $\fbm$ dependence of $\Ctissue$ via Eq.~\eqref{cmacro}, and a method to estimate the strains and stresses in the bone matrix at the microscopic level from those known at the tissue level: 
\begin{align}
    \b \varepsilon^\micro_\bm(\b r, t) &= \mathbb{A}_\bm(\fbm) : \b \varepsilon^\tissue \label{micro-macro-strain}
    \\\b \sigma^\micro_\bm(\b r, t) &= \cmicro_\bm : \big(\mathbb{A}_\bm(\fbm) : \b \varepsilon^\tissue\big) \nonumber
    \\ &\equiv \mathbb{B}_\bm(\fbm): \b \sigma^\tissue,\label{micro-macro-stress}
\end{align}
where Hooke's law~\eqref{hooke} was used in the last equality in Eq.~\eqref{micro-macro-stress}. The stiffness tensor $\Ctissue(\fbm)$, the strain concentration tensor $\mathbb{A}_\bm(\fbm)$, and the stress concentration tensor $\mathbb{B}_\bm(\fbm)$ can be evaluated numerically at each location $\b r$ in the femur midshaft cross section and each time $t$ based on the value of $\fbm(\b r, t)$ and the expressions given in \cite{Fritsch2007} and~\cite{Scheiner2013}. 

Combined with beam theory, this procedure enables us to completely determine, at each time $t$, the spatial distribution across the femur midshaft of (i) the tissue-scale stress and strain tensors, $\b \sigma^\tissue$, $\b \varepsilon^\tissue$; and (ii) the microscopic stress and strain tensors of bone matrix, $\b \sigma_\bm^\micro$, $\b \varepsilon_\bm^\micro$.

In this paper we will consider both the tissue-scale strain energy density (SED), $\sedtissue$, ~and microscopic SED of the bone matrix, $\sedmicro$, as local mechanical quantities sensed and transduced by bone cells. These SEDs are defined by:
\begin{align}
    &\sedtissue(\b r, t) = \tfrac{1}{2} \b \varepsilon^\tissue : \Ctissue : \b \varepsilon^\tissue,\label{macro-sed} \\
    &\sedmicro(\b r, t) = \tfrac{1}{2} \b \varepsilon_\bm^\micro : \cmicro_\bm : \b \varepsilon_\bm^\micro. \label{micro-sed}
\end{align}
The SEDs defined in Eqs \eqref{macro-sed} and \eqref{micro-sed} will be used to formulate biomechanical regulation in the bone remodelling equations. In the literature, biomechanical regulation is commonly based on the SED since this quantity is scalar and it integrates both microstructural state and loading environment [\cite{Fyhrie1986, Mullender1994, Ruimerman2005}]. 

\subsection{Bone tissue remodelling}
The tissue is assumed to be remodelled by a population of active osteoclasts (\oca) and active osteoblasts (\oba). Active osteoclasts are assumed to resorb bone at rate \kres\ (volume of bone resorbed per cell per unit time). Active osteoblasts are assumed to secrete new bone matrix at rate \kform\ (volume of bone formed per cell per unit time). These cellular resorption and formation rates are taken to be constant and uniform. However, the bone volume fraction $\fbm(\b r, t)$ of the tissue may evolve with site-specific rates depending on the balance between the populations of active osteoclasts and active osteoblasts~[\cite{Martin1972,Buenzli2013}]:
\begin{align}\label{fbm}
    \tpd{}{t}\fbm(\b r, t) = \kform \oba - \kres \oca.
\end{align}
In Eq.~\eqref{fbm}, $\oca(\b r, t)$ and $\oba(\b r, t)$ denote the average densities of active osteoclasts and active osteoblasts in the tissue located at $\b r$ (number of cells in the \rve /\tv, Figure \ref{multiscale}(d)). The site-specific remodelling rate $\chi_\bv(\b r, t)$ of the tissue at $\b r$ (also called turnover rate) can be defined as the volume fraction of bone in the RVE that is resorbed and refilled in matched amount per unit time [\cite{Parfitt1983}, Sec.~II.C.2.c.ii]. This corresponds to the minimum of the volume fraction of bone resorbed per unit time, $\kres \oca$, and volume fraction of bone formed per unit time, $\kform \oba$:
\begin{align}\label{turnover}
    \chi_\bv(\b r, t) = \min\{\kres \oca, \kform \oba\}.
\end{align}
Any imbalance between resorption and formation in Eq.~\eqref{fbm} is interpreted as surplus resorption or surplus formation with respect to the baseline of bone properly turned over in Eq. \eqref{turnover}.

Equation~\eqref{fbm} enables us to track site-specific modifications of the midshaft tissue microstructure through $\fbm(\b r, t)$, from which stress and strain distributions across the midshaft can be estimated at both the tissue scale and the microscopic, cellular scale, by means of Eqs~\eqref{beam_theory_strain}--\eqref{beam_theory_stress} and \eqref{micro-macro-strain}--\eqref{micro-sed}.

\subsection{Bone cell population model}\label{section_bone_cell_pop_methods}
\begin{figure*}[t!]
\begin{center}
\includegraphics[width = 0.8\textwidth]{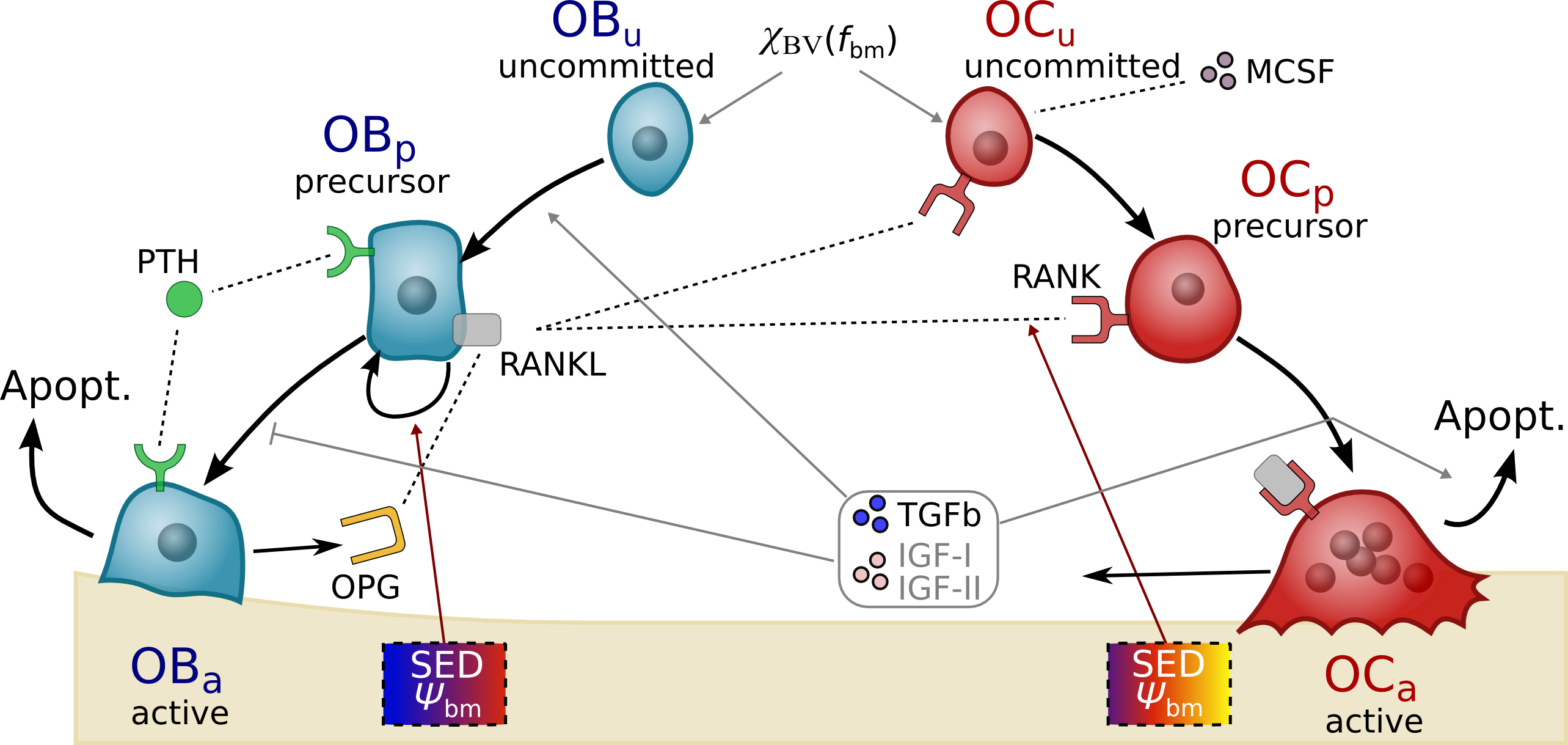}
\caption{Proposed cell population model of bone remodelling taking into account several developmental stages of osteoblasts and osteoclasts together with biochemical regulation, biomechanical regulation (via strain energy density, $\Psi$) and geometrical regulation (via the turnover function, $\chi_{\bv}(\fbm)$).}
\label{cell_pop_model}
\end{center}
\end{figure*}

It remains to specify how the populations of active osteoclasts $\oca(\b r, t)$ and active osteoblasts $\oba(\b r, t)$ evolve in the RVE located at $\b r$ under mechanobiological, geometrical and biochemical regulations. For this, we use a continuum cell population model based on rate equations, originally developed by \cite{Lemaire2004}, and later refined and extended by Pivonka and co-workers [\cite{Pivonka2008, Pivonka2010a, Buenzli2013a, Pivonka2013, Scheiner2013, Pivonka2012}].

To highlight important biochemical couplings and regulations in osteoclastogenesis and osteoblastogenesis, several differentiation stages of osteoclasts and osteoblasts are considered. These biochemical interactions are mediated by several signalling molecules whose binding kinetics are explicitly considered in the model, such as transforming growth factor $\betaup$ (\tgfb), receptor--activator nuclear factor $\kappa$B (\rank) and associated ligand \rankl, osteoprotegerin (\opg), and parathyroid hormone (\pth). The biochemical network of these couplings and regulations is summarised in Figure~\ref{cell_pop_model}. 

Active osteoclasts (\oca s) denote cells attached to the bone surface that actively resorb bone matrix. These cells are assumed to differentiate from a pool of osteoclast precursor cells (\ocp s) by the binding of \rankl\ to the \rank\ receptor, expressed on \ocp s, which induces intracellular \nfkb\ signalling. Osteoclast precursors are assumed to differentiate from a pool of uncommitted osteoclasts progenitors (\ocu), such as haematopoietic stem cells, under the action of macrophage colony stimulating factor (\mcsf) and \rankl\ signalling [\cite{Roodman1999, Martin2004}].

Active osteoblasts (\oba s) denote cells at the bone surface that actively deposit new bone matrix. These cells are assumed to differentiate from a pool of osteoblast precursor cells (\obp s). This activation is inhibited in the presence of \tgfb. Osteoblast precursors are assumed to differentiate from a pool of uncommitted osteoblasts progenitors (\obu), such as mesenchymal stem cells or bone marrow stromal cells, upon \tgfb\ signalling [\cite{Roodman1999}].

The rate equations governing the evolution of the tissue-average cell densities are given by:
\begin{align}
\tpd{}{t}\ocp(\b r, t) = &\docu\big(\mcsf, \rankl(\Psi,\pth)\big)\, \ocu(\fbm) \nonumber \\
&- \docp\big(\rankl(\Psi,\pth)\big)\, \ocp \label{dOCpdt},
\\\tpd{}{t}\oca(\b r, t) = &\docp\big(\rankl(\Psi,\pth)\big)\, \ocp \nonumber \\
&- \aoca(\tgfb)\, \oca \label{dOCadt},
\\\tpd{}{t}\obp(\b r, t) = &\dobu(\tgfb)\, \obu(\fbm) + \pobp(\Psi) \, \obp \nonumber \\
&- \dobp(\tgfb)\, \obp \label{dOBpdt},
\\\tpd{}{t}\oba(\b r, t) = &\dobp(\tgfb)\, \obp \nonumber \\
&- A_\oba \oba \label{dOBadt},
\end{align}
where $\mathcal{D}_i$ is the differentiation rate of cell type $i$ ($i=\ocu, \ocp, \obu, \obp$) modulated by signalling molecules, $\aoca$ is the apoptosis rate of active osteoclasts modulated by \tgfb, $A_\oba$ is the (constant) apoptosis rate of active osteoblasts, $\pobp$ is the proliferation rate of osteoblast precursor cells, and $\Psi$ is the strain energy density, taken to be either $\Psi^\tissue$ or $\Psi^\micro_\bm$. 

The concentrations of the signalling molecules are governed by rate equations expressing mass action kinetics of receptor--ligand binding reactions. Since time scales involved in cell differentiation and apoptosis are much longer than characteristic times of receptor--ligand binding reactions, the signalling molecule concentrations can be solved for in a quasi-steady state (adiabatic approximation) [\cite{Buenzli2013a, Pivonka2012}]. 

Explicit expressions for the signalling molecules concentrations and their modulation of the cell differentiation and apoptosis rates depending on receptor--ligand binding are presented in Appendix \ref{appendix_molecule_formulation} and \ref{appendix_calib}. Below, we comment in more detail on new features of Eqs~\eqref{dOCpdt}--\eqref{dOBadt} that are included to model the geometrical and biomechanical feedbacks on bone cell populations.

\subsubsection*{Geometrical feedback and turnover rate}
\begin{figure*}[t!]
\begin{center}
\includegraphics[trim = 0cm 21cm 0cm 0cm, clip=true, width=0.85\textwidth]{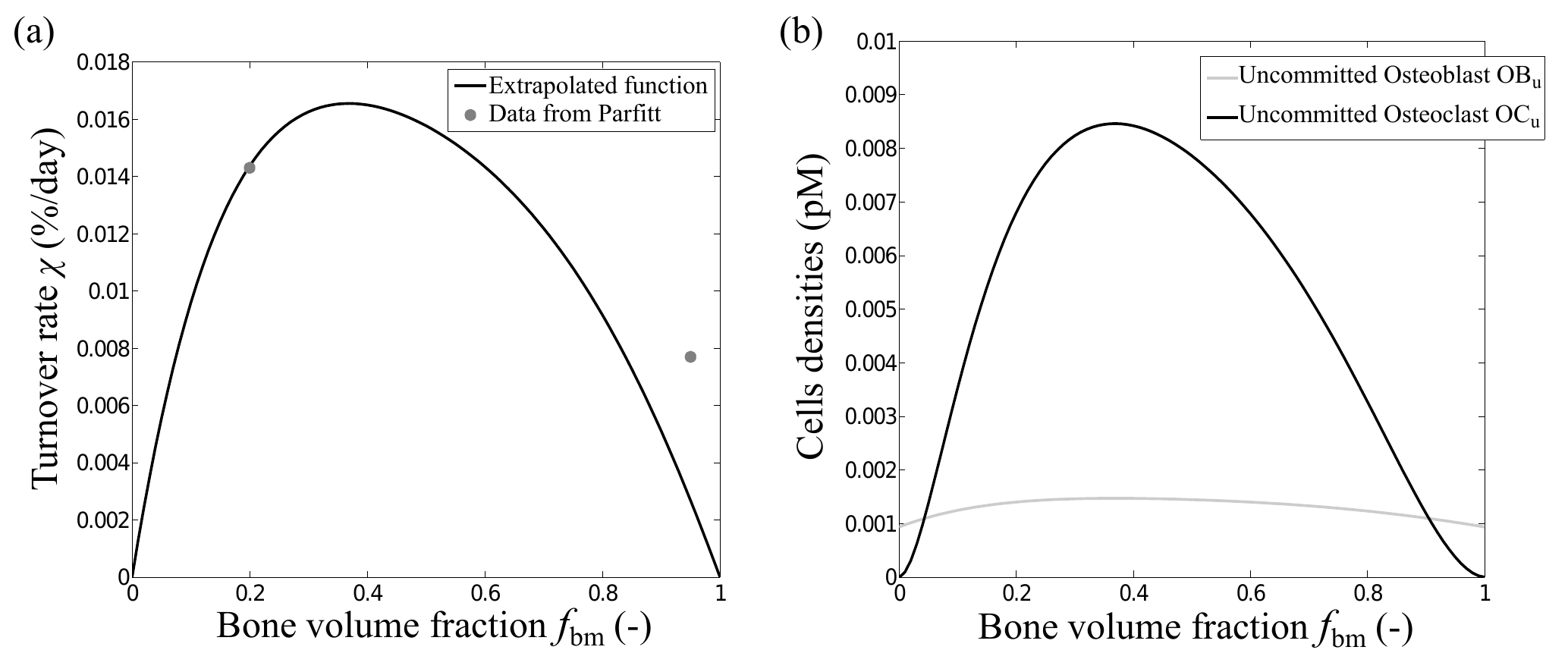}
\caption{(a) Plot of the phenomenological relationship $\chi_\bv(\fbm)$ between turnover rate and bone volume fraction assumed in the model, the grey data points are the ones given by Parfitt [\cite{Parfitt1983}]. (b) Dependence of $\ocu$ and $\obu$ upon $\fbm$ assumed in the bone cell population model.}
\label{turnover_functions}
\end{center}
\end{figure*}

The local availability of bone surface to osteoclasts and osteoblasts is an important factor determining the propensity of initiating new bone remodelling events~[\cite{Martin1972, Buenzli2013}]. A remarkable relationship between the density of bone surface, $\sv$, and bone volume fraction, \fbm, has been exhibited in bone tissues across wide ranges of porosities~[\cite{Martin1984, Fyhrie1999, Lerebours2014}]. This property is particularly interesting from a computational modelling perspective as it enables to track microstructural changes of bone tissues through the evolution of bone volume fraction only. 

In femur midshafts, bone tissue is usually compact, the bone volume fraction is high. However, during bone loss, bone volume fraction tends to decrease in the endocortical region. Due to the fact that \fbm ~reaches values similar to trabecular bone volume fractions, this tissue has been called ``trabecularised" cortical bone [\cite{Zebaze2010}]. Here, we treat compact and porous tissues differently in terms of bone turnover rates [\cite{Parfitt1983, Martin1998}]. Different turnover rates in Eq. \eqref{turnover} can be achieved by assuming that \obu ~and \ocu ~are functions of the bone matrix volume fraction, \fbm, which introduces a dependence of the active bone cell populations, $\oca$ and $\oba$, upon $\fbm$ via Eqs. \eqref{dOCpdt}--\eqref{dOBadt}. This dependence may account both for the influence of bone surface availability on turnover rate via the $\sv(\fbm)$ relation, and for influences of the biochemical microenvironment between cortical bone and trabecularised bone. 

Few experimental data explicitly associate turnover rate with microstructure. However, since the relationship $\sv(\fbm)$ is well established experimentally, it can be expected that a phenomenological relationship, $\chi_\bv(\fbm)$, associating bone turnover with bone volume fraction is well-defined. Parfitt reports that cortical bone of average bone volume fraction 0.95 has a turnover rate of 0.115 $\cm^3/\da$, corresponding to $\chi_\bv(0.95) \approx  \num{0.77}\cdot 10^{-4}/\da$ with $\tv^{\textrm{cort}} = 1.5 \cdot 10^{6} ~\mm^{3}$. Moreover, he states that trabecular bone of average bone volume fraction 0.20 has a turnover rate of 0.25 $\cm^3/\da$, corresponding to $\chi_\bv(0.20) \approx \num{1.43e-4}/\da$ with $\tv^{\textrm{trab}} = 1.75 \cdot 10^{6} ~\mm^{3}$ ~[\cite{Parfitt1983}, Table 1 and Table 7]. We take for $\chi_\bv(\fbm)$ a dome-shaped function following Parfitt's reported values and having a zero turnover rate for \fbm ~equal to 0 and 1 (Figure~\ref{turnover_functions}(a)). The maximum of bone turnover is assumed to occur at \fbm = 0.35, corresponding to typical trabecular or trabecularised bone microstructures. These types of microstructures are expected to remodel at the highest rates due to the proximity of precursor cells in the marrow and the large availability of bone surface.

The functions $\ocu(\fbm)$ and $\obu(\fbm)$ are determined such that the turnover rate obtained from the cell population model in a normal healthy state with balanced remodelling, matches the phenomenological relationship $\chi_\bv(\fbm)$. From Eqs~\eqref{fbm} and~\eqref{turnover}, the balanced steady-state condition and the remodelling rate condition impose the constraints
\begin{align}
\chi_\bv(\fbm) &= \kform \overline\oba\big(\ocu(\fbm), \obu(\fbm)\big)  \label{turnover_OBa}\\
&= \kres \overline\oca\big(\ocu(\fbm), \obu(\fbm)\big) \label{turnover_OCa}
\end{align}
at each value of $\fbm\in [0,1]$, where the bar indicates steady-state values of the cell density variables. These two constraints were solved numerically with the turnover rate function $\chi_\bv(\fbm)$ reported in Figure \ref{turnover_functions}(a) by using a trust-region dogleg method. The functions $\ocu(\fbm)$ and $\obu(\fbm)$ obtained by this procedure are shown in Figure \ref{turnover_functions}(b). These functions are used as input in Eqs \eqref{dOCpdt}--\eqref{dOBadt}  in all our simulations. This ensures that in steady state, each \rve ~of the midshaft cross section located at $\b r$ remodels at rate $\chi_\bv(\fbm(\b r))$ without changing its bone volume fraction. The functions $\ocu(\fbm)$ and $\obu(\fbm)$ are assumed to hold unaffected in the various deregulating circumstances considered later on (i.e., osteoporosis and altered mechanical loading).

The explicit calibration of the cell population model, Eqs \eqref{dOCpdt}--\eqref{dOBadt}, to site-specific tissue remodelling rates is a significant novelty compared to our previous temporal model [\cite{Pivonka2013, Scheiner2013}]. This modification was made necessary to consistently describe the site-specific evolution of bone in the spatio-temporal framework of ~\cite{Buenzli2013} whilst retaining a cell population model that includes biochemical regulations.

\subsubsection*{Mechanical feedback and initial bone microstructure stability}
A mechanical feedback is included in the cell population model such that underloaded regions of bone promote osteoclastogenesis and overloaded regions of bone promote osteoblastogenesis [\cite{Frost1987, Frost2003}]. These responses are viewed as consequences of biochemical signals transduced from a mechanical stimulus sensed by osteocytes [\cite{Bonewald2011}]. Osteocytes are known to express \rankl, which regulates osteoclast generation, and sclerostin, which regulates osteoblast generation via \wnt\ signalling [\cite{Bonewald2008}]. Following ~\cite{Scheiner2013}, the resorptive response of the mechanical feedback is assumed to act by an increase in the microenvironmental concentration of \rankl, whereas its formative response is assumed to act by an increase in the proliferation rate of osteoblast precursors. 

The exact nature of the mechanical stimulus sensed by osteocytes is still a matter of debate. It may include lacuno-canalicular extracellular fluid shear stress on the osteocyte cell membrane, extracellular fluid pressure, streaming potentials and direct deformations of the osteocyte body induced by bone matrix strains [\cite{KnotheTate2003, Bonewald2008, Bonewald2011}]. Due to the extensive network of osteocyte connections in bone~[\cite{Buenzli2015}], average bone matrix strains at a higher scale may also be sensed by the osteocyte network. Below, we assume that the mechanical stimulus to the bone cell population model is described by a local strain energy density, $\Psi(\b r, t)$. This strain energy density will be taken to be either the microscopic, cellular-scale strain energy density of bone matrix, $\Psi^\micro_\bm(\b r, t)$, or the average tissue-scale strain energy density, $\Psi^\tissue(\b r, t)$, defined respectively in Eqs~\eqref{micro-sed} and~\eqref{macro-sed}.

To predict with our model the evolution of a real scan of midshaft femur under various deregulating circumstances, it is important to assume that the bone scan represents a stable state initially in absence of any deregulation. In particular, this initial bone state is assumed mechanically optimal. This can be ensured by choosing the local mechanical stimulus acting onto the bone cells, $\mu(\b r, t)$, as a normalised difference between the current SED and the SED of the inital bone microstructure $\Psi(\b r, 0)$:
\begin{align}\label{mech-stimulus}
\mu(\b r, t) = \frac{\Psi(\b r, t) - \Psi(\b r, 0)}{\Psi(\b r, 0) + K}
\end{align}
The normalisation by $\Psi(\b r, 0)$ in the denominator in Eq.~\eqref{mech-stimulus} ensures that the mechanical stimulus is not over-emphasised away from the neutral axis where strain energy density takes high values. The small positive constant $K=1 \cdot 10^{-6} ~\textrm{GPa}$ is added to keep mechanical stimulus well defined near the neutral axis where $\Psi(\b r, 0) \approx 0$ (see also Discussion section \ref{discussion_mecha}). 

When negative, $\mu(\b r, t)$ in Eq.~\eqref{mech-stimulus} is assumed to promote $\beta_\rankl^\text{mech}$, the production rate of \rankl:
\begin{align}\label{kappa_eq}
    \beta_\rankl^\text{mech}(\Psi) = \begin{cases}
     - \kappa \cdot \mu(\b r, t), &\quad \text{if } \mu(\b r, t) \leq 0
    \\0, &\quad \text{if } \mu(\b r, t) > 0
    \end{cases}
\end{align}
where $\kappa$ is a parameter describing the strength of the biomechanical transduction (see section \ref{Calibration_mecha_section}). This results in increased \rankl\ signalling in underloaded conditions (see Eq.~\eqref{RANKL} in Appendix), and so in increased osteoclast generation in Eqs~\eqref{dOCpdt}--\eqref{dOCadt}.

When positive, $\mu(\b r, t)$ in Eq.~\eqref{mech-stimulus} is assumed to promote $\pobp$, the proliferation rate of pre-osteoblasts in Eq.~\eqref{dOBpdt}: 
\begin{align}\label{proliferation_eq}
    \pobp(\Psi) = P_\obp + 
        \begin{cases}
        0, &\text{if } \mu(\b r, t) \leq 0
        \\P_\obp \cdot \lambda \cdot \mu(\b r, t), &\text{if } 0 < \mu(\b r, t) < \frac{1}{\lambda}
        \\P_\obp, &\text{if } \mu(\b r, t) \geq \frac{1}{\lambda}
        \end{cases}
\end{align}
where $\lambda$ is a parameter describing the strength of the biomechanical transduction. The first term in \eqref{proliferation_eq} accounts for a transit-amplifying stage of osteoblast differentiation occurring in absence of mechanical stimulation [\cite{Buenzli2013a}]. The proliferation rate is assumed to saturate to the value $\pobp = 2 P_\obp$ in highly overloaded situations to ensure the stability of the population of \obp s [\cite{Buenzli2013a, Scheiner2013}]. 

A similar type of mechanical feedback was implemented in purely temporal settings in \cite{Scheiner2013}. The initial strain energy density distribution, $\Psi(\b r, 0)$, is calculated from Eqs~\eqref{macro-sed}--\eqref{micro-sed} and from the initial bone volume fraction distribution, $\fbm(\b r, 0)$, determined on the bone scan (described in the next section).

\subsection{Initial distribution of bone volume fraction from microradiographs}\label{microCT_section}
\begin{figure*}[t!]
\begin{center}
\includegraphics[trim = 0cm 8.5cm 0cm 0cm, clip=true,  width = 0.9\textwidth]{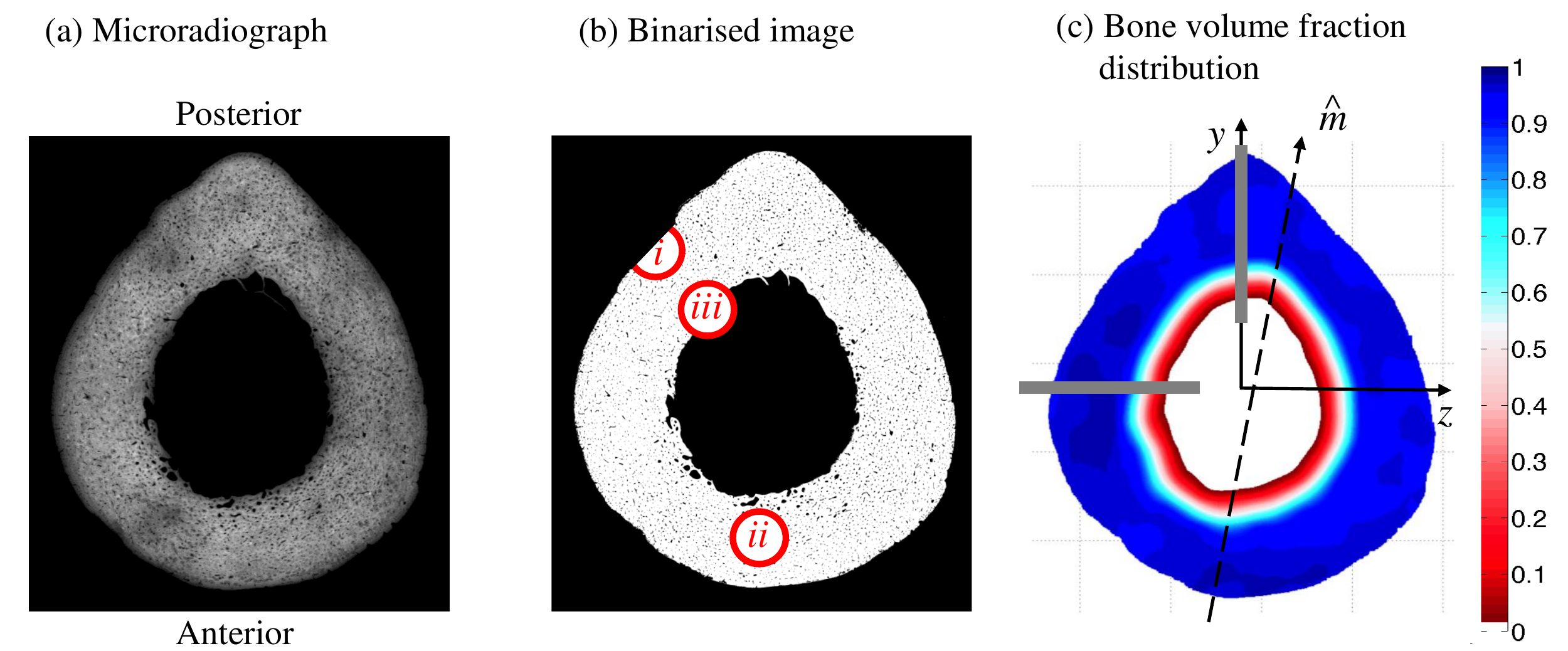}
\caption{(a) Microradiograph of a midshaft femur cross section (courtesy of C. David L. Thomas and John G. Clement, Melbourne Femur Collection). (b) Binarisation of the microradiograph and determination of the local \fbm ~values: (i) at the periosteal region, (ii) in the intracortical region, and (iii) at the endosteal region; (c) Bone volume fraction distribution extracted from the radiograph and interpolated. The dashed line represents the location of the neutral axis. The origin of the coordinate system $(y, z)$ is taken at the normal force center, NC. The grey lines are the 10 mm along which we are studying the evolution of the model in the Results Section.}
\label{extraction}
\end{center}
\end{figure*}

The initial microstructural state of the midshaft bone cross section can be derived from high-resolution bone scans such as micro-computed tomography (microCT) scans or microradiographs. Since Haversian canals have an average diameter of about 40\,$\mu$m, at least 10\,$\mu$m\ resolution is required to evaluate intracortical bone volume fractions with sufficient accuracy. 

In the simulations presented in Section~\ref{results}, we used the microradiograph represented in Figure~\ref{extraction}(a) where the pixel size is 7 $\mu$m. The femur sample was collected from a 21-year-old subject. The microradiograph was digitised and binarised by a thresholding operation based on pixel grey level. Bone matrix is assigned the value 1 irrespective of the degree of mineralisation, and intracortical pores are assigned the value 0. The distribution of the bone volume fraction, $\fbm(\b r, 0)$, across the midshaft was determined by calculating the volume of bone matrix in a disk of 2 mm diameter, centred at each pixel of the binarised image and divided by the disk's area. For the points near the periosteal surface, only the portion of the disk contained into the subperiosteal area was used for this calculation (see Figure~\ref{extraction}(b)). The discrete values of $\fbm$ defined at each pixel contained in the subperiosteal region were then interpolated into a continuous function, $\fbm(\b r, 0)$, using \texttt{Matlab}'s 2D cubic interpolation procedure. The result is shown in Figure~\ref{extraction}(c). A similar exclusion was not performed at the endosteal surface since this surface is less well defined, in opposition to the periosteal surface, due to the presence of `trabecular-like structures. Bone matrix volume fractions near the endosteal surface are averages of intracortical bone regions and regions in the bone marrow cavity.

\begin{figure*}[t!]
\begin{center}
\includegraphics[trim = 0cm 20cm 0.5cm 0.9cm, clip=true,  width=0.85\textwidth]{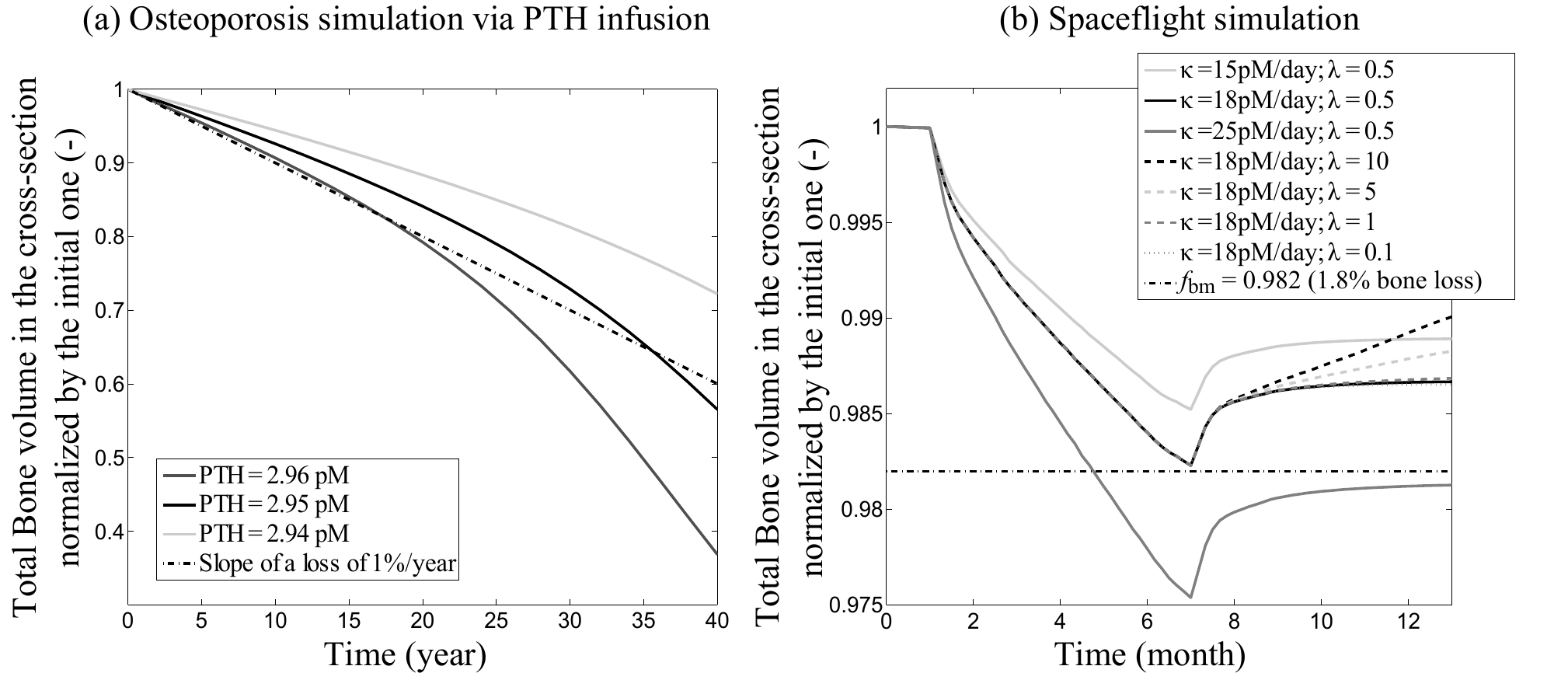}
\caption{(a) Evolution of the total bone mass in the cross section with time while simulating osteoporosis: calibration of the PTH infusion. Osteoporosis is characterised by a bone loss of 1\%/year [\cite{Parfitt1987b, Nordin1988, Szulc2006}]. (b) Evolution of the total bone mass in the cross section with time while simulating a spaceflight mission: calibration of the mechanical parameters, $\lambda$ and $\kappa$.}
\label{Calibration}
\end{center}
\end{figure*}

\subsection{Numerical simulations}
The multiscale mechanobiological model of bone remodelling presented in this paper is governed by a coupled system of (i) distributed ODEs describing the evolution of bone cell populations at each location $\b r$ in the midshaft femur (Eqs (\ref{dOCpdt})--(\ref{dOBadt})); and (ii) non-local and tensorial algebraic equations determining the mechanical state of the tissue \rve ~at $\b r$, both at the tissue scale and at the microscopic scale (Eqs \eqref{Bernoulli}--\eqref{micro-sed}). The model is initialised with a bone volume fraction distribution across the midshaft femur deduced from high-resolution bone scans (Figure \ref{extraction}(a)) and with steady-state populations of cells fulfilling the site-specific turnover rate conditions Eqs. \eqref{turnover_OBa}--\eqref{turnover_OCa}. This initial state is thereby constructed to be a steady state of the model, in which the biochemical, geometrical and mechanobiological regulations of resorption and formation are balanced. 

To solve this non-local spatio-temporal problem numerically, we use a staggered iteration scheme in which we first solve the mechanical problem (i.e., tissue-scale SED and microscopic SED) assuming constant material properties, and then solve the bone cell population model and evolve the bone volume fraction at each location $\b r$ of the femur midshaft assuming constant mechanical feedback for a duration $\Delta$t. After $\Delta$t, the mechanical problem is recalculated based on the updated bone volume fraction distribution, \fbm($\b r$, t+$\Delta$t), and this procedure is iterated. The ODEs are solved using a standard stiff ODE solver (\texttt{Matlab, ode15s}). The spatial discretisation is a regular grid with steps $\Delta y = \Delta z = 0.8 ~\textrm{mm}$. Due to the separation of time scales between changes in the local mechanical environment (years) and changes in bone cell populations (days), the mechanical stimulus requires updating after durations $\Delta$t = 2 years. The accuracy of the numerical results depending on $\Delta$t is studied in Appendix \ref{appendix_delta_t}.

\subsection{Model calibration}
The model presented in this paper contains: (i) biomechanical parameters associated with the estimation of $\Psi(\b r,t)$, and (ii) parameters associated with the bone cell population model. Biomechanical parameters as well as biochemical parameters were determined and validated in other studies [\cite{Scheiner2013, Pivonka2013, Pivonka2008, Buenzli2013a}] (Table \ref{nomenclature}). Here we calibrate the newly introduced parameters: (a) mechanical coupling parameters $\lambda$ and $\kappa$ (Eqs \eqref{kappa_eq} and \eqref{proliferation_eq}), and (b) biochemical parameters related to the simulation of osteoporosis. 

\subsubsection*{Calibration of the hormonal deregulation for osteoporosis simulation}
In our previous temporal model [\cite{Scheiner2013}], osteoporosis was modelled by an increase in systemic \pth ~together with a reduction in the biomechanical transduction parameters: $\lambda$ and $\kappa$. In this paper, we simulate age-related bone loss using a single parameter perturbation, i.e., an increase in systemic \pth ~concentration. This increase is calibrated so as to obtain a loss of total bone cross-sectional area in the femur midshaft of 1\% per year \footnote{The calibration is performed without mechanical adaptation (i.e. setting $\lambda$ = 0 and $\kappa$ = 0 in Eqs \eqref{kappa_eq} and \eqref{proliferation_eq}) in order to compare both mechanical feedbacks in a more consistent way.} [\cite{Parfitt1987b, Nordin1988, Szulc2006}]. The total bone cross-sectional area is defined by the integral of $\fbm(\b r, t)$ over the midshaft cross section. In the model, a rate of bone loss of 1\%/year was obtained by an increase in systemic concentration of \pth ~from 2.907 pM to 2.954 pM (1.62\% increase) (see Figure \ref{Calibration}(a)).

\subsubsection*{Calibration of mechanobiological feedback}\label{Calibration_mecha_section}
The rate of change in bone mass due to mechanical feedback is determined in the model by the biomechanical transduction parameters $\lambda$ (in Eq. \eqref{proliferation_eq}) and $\kappa$ (in Eq. \eqref{kappa_eq}). To calibrate these parameters, we used data gathered from mechanical disuse and re-use experiments. It has been shown that cosmonauts undertaking long mission space flights lose bone mass at a rate of approximately 0.3\% per month [\cite{Vico2000}]. This microgravity-induced bone loss is only slowly recovered after return to Earth. No significant bone gain is observed after 6 month exposure to normal gravity on Earth [\cite{Vico2000, Collet1997}]. 

In our multiscale model, microgravity is simulated as a 80\% reduction of the normal mechanical loads experienced by the femur, i.e., $\b N_{\textrm{microgravity}} = 0.2 \b N$ and $\b M_{\textrm{microgravity}} = 0.2 \b M$. Based on these reduced loads, the parameter $\kappa$ has been calibrated such that $1.8\%$ of total bone cross-sectional area is lost after 6 months. We found such rate of loss with $\kappa = 18$ pM/day when the mechanical stimulus is based on the microscopic SED, $\Psi^\micro_\bm(\b r, t)$ (see Figure \ref{Calibration}(b)), and $\kappa = 19$ pM/day when the mechanical stimulus is based on the tissue-scale SED, $\Psi^\tissue(\b r, t)$. After return to Earth, rates of bone recovery are too low to be detected after 6 months [\cite{Collet1997}]. We performed a parametric study investigating various strengths of $\lambda$. Using parameter values of $\lambda > 1$ in our model results in significant bone gain after 6 months, while $\lambda \leq 1$ results in small bone gain. Based on these results we use $\lambda = 0.5$ for both the microscopic and tissue-scale mechanical stimuli.

\section{Results}\label{results}
\begin{figure*}[t!]
\begin{center}
\includegraphics[width=0.8\textwidth]{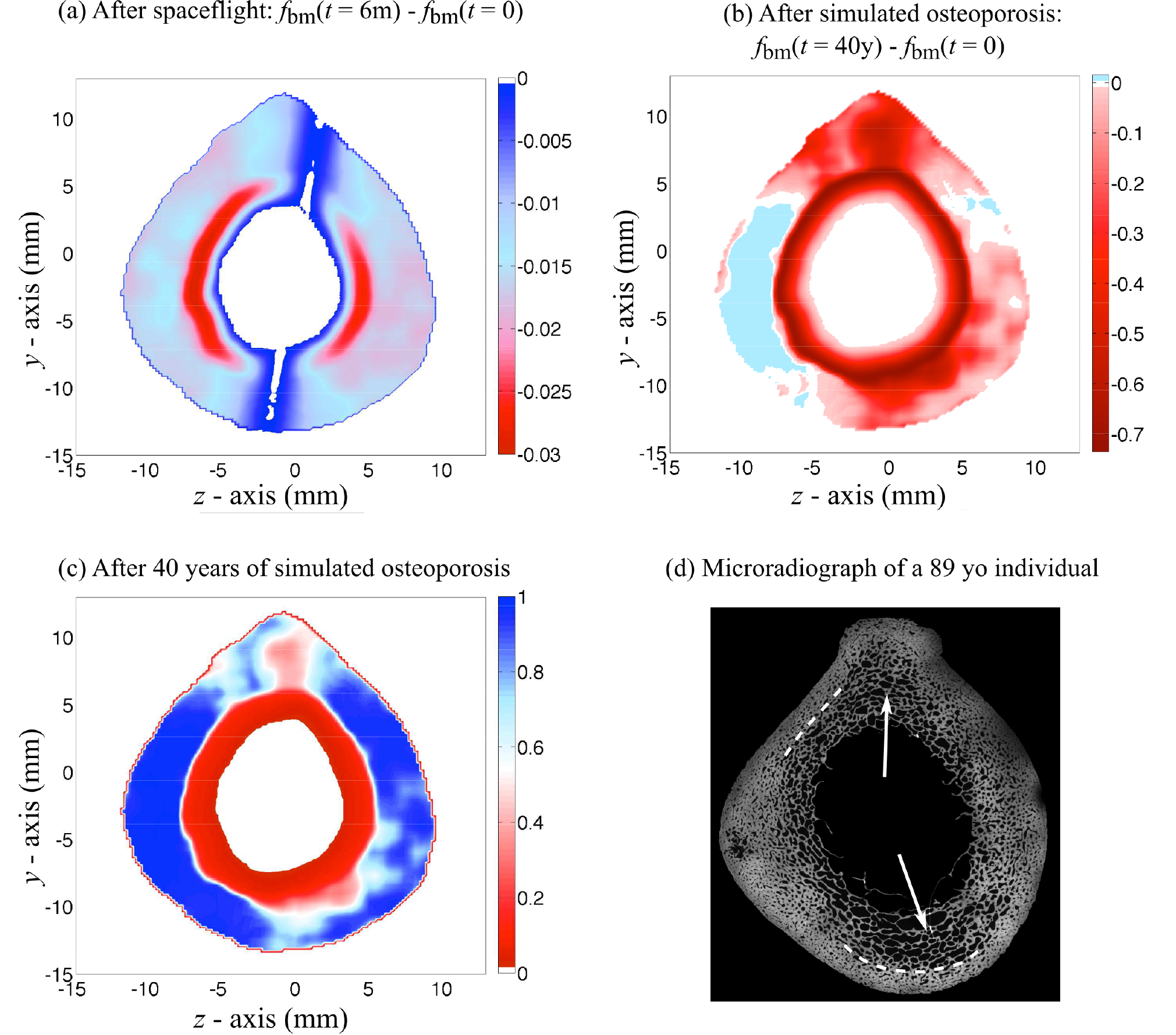}
\caption{(a) Difference between the bone volume fraction distribution in the cross section after a 6-months spaceflight mission with the initial distribution. (b) Difference between the bone volume fraction distribution in the cross section after a 40-years of simulated osteoporosis with the initial distribution. (c) Bone Volume Fraction distribution in the cross section after 40 years of simulated osteoporosis. (d) Microradiograph of a human femur cross section from an 89 years old individual. The dashed lines highlight regions with sharp transition between porous and compact tissue. The arrows point out regions with high porosity along the antero-posterior axis.}
\label{Model_check}
\end{center}
\end{figure*} 

In this section we present numerical simulations of the evolution of the midshaft femur cross section subjected to either: (i) changes in mechanical environment (Section \ref{results_mecha_disuse}), or (ii) hormonal deregulation simulating osteoporosis (Section \ref{results_OP}). We also investigate how site-specific bone loss may depend on whether mechanical stimulus is sensed at the microscopic, cellular scale, or at the tissue scale.

\subsection{Bone loss due to mechanical disuse}\label{results_mecha_disuse}
Figure \ref{Model_check}(a) represents site-specific changes of the femur midshaft cross section simulated by the model assuming a 80\% reduction in the normal mechanical loading. This reduction in mechanical loading may represent microgravity in long spaceflight missions (see Section \ref{Calibration_mecha_section}) or prolonged bed rest. The Figure depicts the difference between the bone volume fraction distribution after 6 months of mechanical disuse and the initial bone volume fraction. It can be seen that bone loss is site-specific with more bone loss occurring near the endosteal surface. Close to the neutral axis, only limited loss of bone is observed.

\subsection{Simulation of osteoporosis due to hormonal deregulation}\label{results_OP}

\begin{figure*}[t!]
\begin{center}
\includegraphics[width=0.8\textwidth]{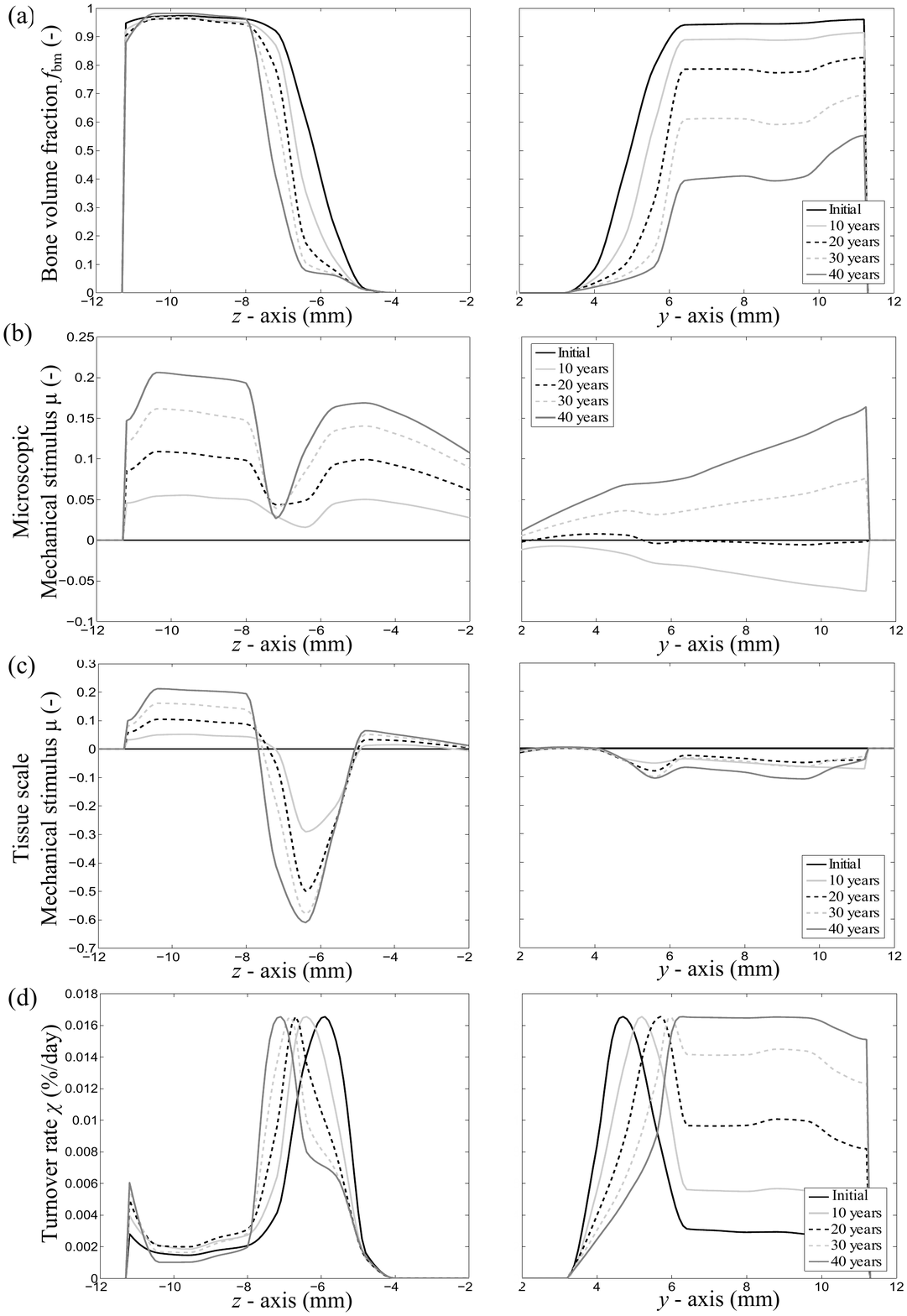}
\caption{Evolution of (a) Bone volume fraction; (b) Mechanical stimulus, $\mu(\b r, t)$, at the microscopic scale; (c) Mechanical stimulus, $\mu(\b r, t)$, at the tissue scale; and (d) Turnover rate along the $y$ and $z$-axis during the simulation of osteoporosis.}
\label{all_plots_mecha}
\end{center}
\end{figure*}

\begin{figure*}[t!]
\begin{center}
\includegraphics[trim = 0.4cm 20.4cm 0cm 0.2cm, clip=true,   width=0.8\textwidth]{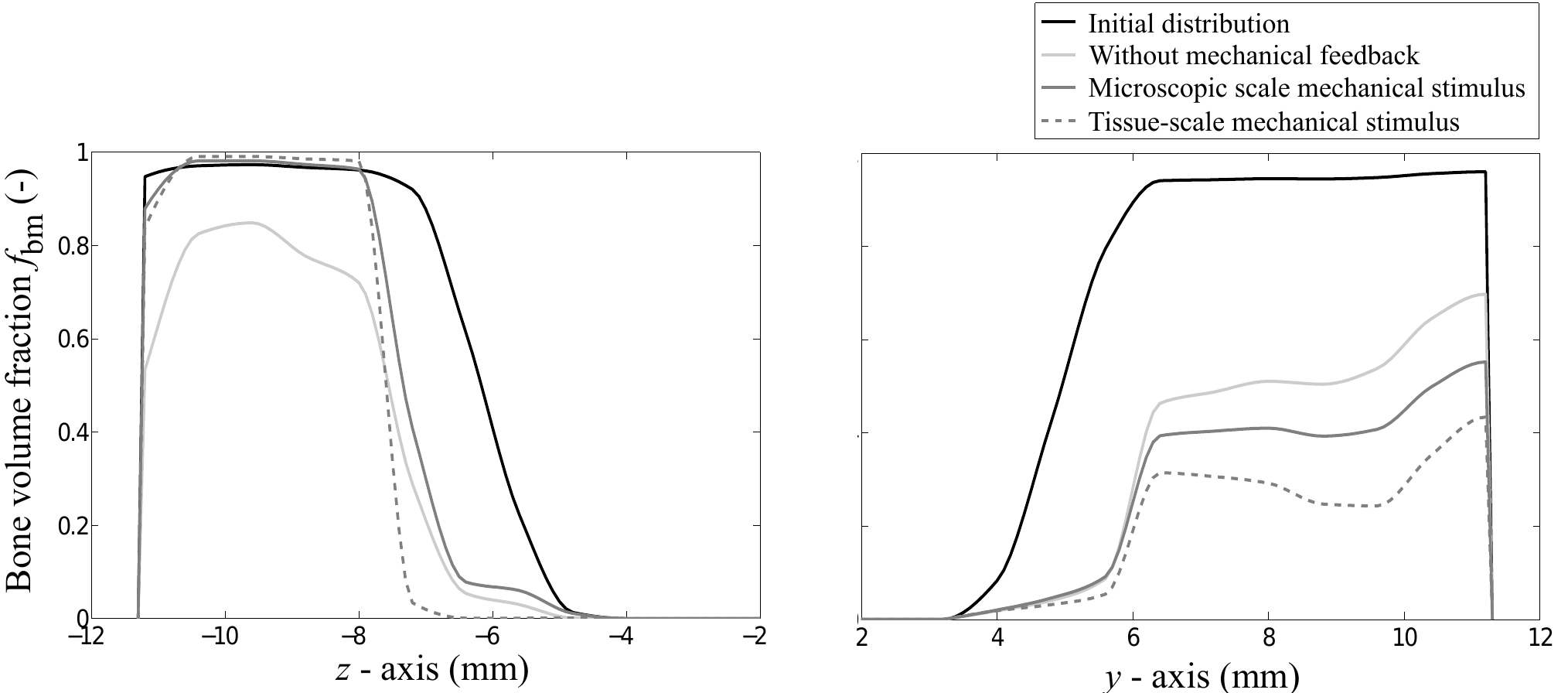}
\caption{Bone volume fraction along the $y$- and $z$-axes, the initial distribution and the distributions after 40 years of simulated osteoporosis, without mechanical regulation, with mechanical regulation based on microscopic SED, and with mechanical regulation based on tissue-scale SED.}
\label{without_with_mecha}
\end{center}
\end{figure*}

Figures \ref{Model_check}(b) and (c) represent the site-specific changes of the midshaft cross section that occur after 40 years of simulated osteoporosis when the mechanical feedback acting onto the bone cell population model is based on the microscopic SED. Figure \ref{Model_check}(b) depicts the difference between the \fbm ~distribution at the end of the simulation and the initial distribution. Figure \ref{Model_check}(c) depicts the \fbm ~distribution at the end of the simulation. Bone loss occurs everywhere in the cross section except at the medial and lateral sides. The loss is site-specific with higher rates of loss in the endocortical region and around the neutral axis, close to the antero-posterior axis. This pattern of bone loss is consistent with the high porosity commonly observed in these regions in osteoporotic subjects (Figure \ref{Model_check}(d), arrows). The simulation exhibits a sharp transition between a very porous endocortical region and a dense intracortical region towards the periosteum. Although perhaps less pronounced, such a transition is also observed in the microradiograph of Figure \ref{Model_check}(d) (dashed lines). In contrast to the osteoporosis simulation, the simulation of mechanical disuse (Figure \ref{Model_check}(a)) shows that bone was lost all over the cross section, with little change around the neutral axis. In both simulations, bone was lost predominantly in the endocortical region. 

In Figure \ref{all_plots_mecha}, we show how the distributions of the following quantities evolved along the $y$- and the $z$-axes during the simulation of osteoporosis: (a) the bone volume fraction, (b) the microscopic mechanical stimulus, $\mu_{\bm}^{\micro}$, used as mechanical stimulus in this simulation, (c) the tissue-scale mechanical stimulus, $\mu^{\tissue}$, not used as mechanical stimulus in this simulation, and (d) the turnover rate. Along both axes, the regions in which bone volume fraction transitions from low to high (3 $< y < $ 6 mm and -7 $< z <$ -5 mm) are resorbed at higher rate, due to the higher values of $\chi_\bv$ in these regions (Figure \ref{all_plots_mecha}(d)). As time progresses, bone volume fraction is strongly reduced in the endocortical region, leading to an expansion of the marrow space and a reduction in cortical wall width. This is accompanied by a shift of the maximum of $\chi_\bv$ towards the periosteum. Along the $y$-axis (near the neutral axis), bone is lost at a high rate not only in the endocortical region but also near the periosteum, as can be seen by the gradual increase in turnover rate in the whole cortical width (Figure \ref{all_plots_mecha}(d)). In contrast, along the $z$-axis, bone is lost at a high rate only at the endosteum where turnover rate maintains a well-defined peak. The intracortical region ($z <$ -7 mm) is preserved even after 40 years of simulated osteoporosis.

\subsubsection*{Microscopic vs tissue-scale mechanical stimulus}
Comparing Figures \ref{all_plots_mecha}(b) and (c), we can observe that the values of the mechanical adaptation stimuli are strongly dependent on the length scale at which they are calculated, i.e. tissue scale or microscopic scale. Along the $z$-axis, $\mu^{\micro}_{\bm}$ is always positive (Figure \ref{all_plots_mecha}(b)), whereas, $\mu^{\tissue}$ takes negative values in the endocortical region (Figure \ref{all_plots_mecha}(c)). Regions with high values of $\mu^{\micro}_{\bm}$ and positive values of $\mu^{\tissue}$ correlate with regions where the bone matrix is preserved. Regions with low values of $\mu^{\micro}_{\bm}$ and negative values of $\mu^{\tissue}$ correlate with regions where the bone matrix is resorbed. The Figures also show a qualitative and quantitative difference in mechanical stimuli $\mu^{\micro}_{\bm}$ and $\mu^{\tissue}$ between the $y$- and $z$-axes. The mechanical stimulus is asymmetric between the antero-posterior axis and lateral-medial axis due to the assumed bending loading state. Along the $y$-axis, no important variation can be observed between endocortical and periosteal regions. Along the $z$-axis, both stimuli exhibit much lower values in the endocortical region than at the periosteum or in the marrow cavity. We note that the mechanical stimulu are not zero in the marrow even when $\fbm = 0$ due to the assumed vascular stiffness.

Figure \ref{without_with_mecha} compares the evolution of bone volume fraction during simulated osteoporosis when the mechanical stimulus acting onto the bone cells is either (i) absent, (ii) based on the microscopic mechanical stimulus, $\mu^{\micro}_{\bm}$, or (iii) based on the tissue-scale mechanical stimulus, $\mu^{\tissue}$.\footnote{For the simulation in case (iii), the mechanical transduction strength parameters are: $\kappa = 19$ pM/day and $\lambda = 0.5$, calibrated with the tissue-scale mechanical stimulus while simulating spaceflight.} All cases exhibit strong endocortical bone loss with little difference in the expansion rate of the medullary cavity. A slightly steeper endosteal wall is created along the $z$-axis during the simulation using tissue-scale mechanical stimulus, and a region with very low bone volume fraction ($\fbm \simeq 0.1$) is preserved in the medullary cavity during the simulation with microscopic mechanical stimulus. Intracortical bone towards the periosteum is preserved along the $z$-axis by both mechanical stimuli, but it is resorbed more strongly along the $y$-axis in the simulation with tissue-scale mechanical stimulus.

\section{Discussion}

\subsubsection*{Endocortical bone loss}\label{discussion_geom}
The loss of endocortical bone, with its associated expansion of the marrow cavity and cortical wall thinning, is a trait common to several bone disorders and deregulations of remodelling. It is observed in osteoporosis [\cite{Feik1997, Parfitt1998, Bousson2001, Thomas2005, Szulc2006, Zebaze2010}], vitamin D deficiency [\cite{Busse2013}], hyperparathyroidism [\cite{Hirano2000, Burr2001, Turner2002}], but also during disruptions of normal mechanical loading, such as in prolonged bed rest [\cite{LeBlanc2007, Rittweger2009}], long term space missions [\cite{Vico2000, Lang2004}], trauma-induced paralysis such as spinal cord injury [\cite{Kiratli2000, Eser2004}], and as well in animal studies: muscle paralysis [\cite{Warner2006, Ausk2012, Ausk2013}] or hind-limb disuse induced by tail suspension [\cite{Bloomfield2002, Judex2004}].

Our numerical simulations of osteoporosis and mechanical disuse are consistent with these experimental findings. All Figures \ref{Model_check}, \ref{all_plots_mecha} and \ref{without_with_mecha} highlight the strong site-specificity of bone loss under deregulations of bone remodelling. The endocortical region systematically undergoes the most significant loss. This similarity arises despite the fact that in the simulation of mechanical disuse, the deregulation is non-uniform in the cross section (due to the uneven distribution of mechanical loads) whereas in the simulation of osteoporosis, the hormonal deregulation is uniform in the cross section.

The precise mechanisms that underlie the predominant loss of bone in the endocortical region are still poorly understood [\cite{Raisz2001, Thomas2005, Squire2008, Ausk2012}]. Mechanical adaptation has been suggested as a potential mechanism [\cite{Frost1997, Burr1997, Thomas2005, Jepsen2012}]. Bone loss induced by mechanical disuse redistributes mechanical loads towards the periosteum, where bone volume fraction is higher. This could unload endocortical regions and thereby accelerate their resorption. Reduced physical activity and muscle strength in ageing subjects support this hypothesis [\cite{Frost1997}]. However, the ubiquity of endocortical bone loss in situations in which mechanical loading is not significantly modified suggests that other mechanisms are at play. The morphological influence of the tissue microstructure on the rate of bone loss has been suggested to be another important factor [\cite{Martin1972, Squire2008, Zebaze2010, Buenzli2013}]. Cortical bone has little bone surface available to bone cells, but this surface expands during bone loss, which could increase the activation frequency of remodelling events. If remodelling is imbalanced, this may lead to an acceleration of bone loss, and to an increase of the available surface until the tissue microstructure becomes so porous that its surface area reduces with further loss [\cite{Martin1972, Raisz2001}]. 

We have shown previously the possibility of this morphological mechanism to explain cortical bone trabecularisation in both temporal [\cite{Pivonka2013}] and spatio-temporal settings [\cite{Buenzli2013}]. The spatio-temporal model proposed in the present work incorporates both mechanical adaptation and a morphological feedback of the microstructure on turnover rate. In Figure \ref{without_with_mecha}, our simulations of osteoporosis conducted with and without mechanical feedback suggest that the rate at which the medullary cavity expands and the cortical wall thins is only marginally dependent on mechanical adaptation. This rate is primarily due to the high turnover rates present in the endocortical region (Figure \ref{all_plots_mecha}(d)), i.e., due to the morphological influence of microstructure on the rate of loss. This proposed mechanism is consistent with the observation that distinct conditions exhibit endocortical bone loss, whether mechanical loading is disrupted or not.

\subsubsection*{Model formulation of morphological feedback}
In the cell population model of \cite{Pivonka2013}, the morphological influence of the tissue microstructure was included through the specific surface of the tissue [\cite{Martin1984, Lerebours2014}] normalised by its initial value. This normalisation allowed to maintain the same cell behaviour in both cortical and trabecular bones. However, it leads to a turnover rate that is initially independent of bone volume fraction, and so the same in cortical and trabecular bone. The morphological feedback proposed in the present model differs by (i) avoiding a dependence on an initial reference state (i.e., absence of normalisation to allow microstructure-dependent turnover rates), and (ii) by referring to turnover rate (a dynamic biological quantity) instead of specific surface (a morphological characterisation of the microstructure).

Whilst specific surface can be estimated directly from high-resolution scans of bone tissues [\cite{Chappard2005, Squire2008, Lerebours2014}], quantitative links between $\sv$ and cell numbers remain unclear [\cite{Martin1972, Parfitt1983, Pivonka2013}]. The direct reference to turnover rate, in the present model, makes the model more accurate, due to the straightforward link between turnover rate and cell populations (see Eq.~\eqref{turnover}). Unfortunately, turnover rate is rarely measured experimentally by cell counts or volumes of bone resorbed and re-formed. It is most commonly characterised by measurements of serum concentrations of bone resorption and/or formation markers [\cite{Szulc2006, Burghardt2010, Malluche2012}], which are difficult to relate quantitatively to cell numbers or bone volume at a particular bone site. Whilst the phenomenological relationship $\chi_\bv(\fbm)$ that we assumed between turnover rate and microstructure remains to be studied quantitatively, such a relationship has been suggested in several studies [\cite{Felsenberg2005, Burghardt2010, Malluche2012}].

\subsubsection*{Nature of the mechanical stimulus}\label{discussion_mecha}
The nature of the mechanical stimulus sensed by bone cells and transduced into signals prompting resorption or formation has been a matter of discussion for many years. A number of computational studies simulating mechanical adaptation of bone microstructure suggested that the strain energy density could be a good candidate. \cite{Ruimerman2005} tested several mechanical stimuli and concluded the SED gave best results when comparing simulations outcomes with biological parameters such as porosity, trabecular number or adaptability to external loading. However, \cite{Levenston1998} argued that the drawback of using the SED is that it does not lead to a different response when bone is stimulated in tension or in compression. In the literature, most computational models use the SED because it is a scalar representing both microstructure and mechanical loading [\cite{Fyhrie1986, Mullender1994, VanRietbergen1999, VanOers2008, Scheiner2013}]. Quantitative criteria based on experimental observations are still lacking, especially ones testing the tensorial aspects of mechanical loading conditions. For our purpose of studying tissue-scale average changes in bone volume fraction, these tensorial aspects are likely to be secondary. Hence, we have based our mechanical stimulus on the strain energy density (see below for a discussion of the scale). We note here that other mechanical quantities have also been proposed and studied for their magnitude and possible influence onto osteocytes, such as fluid shear stress and fluid pressure in the lacuno-canalicular system [\cite{KnotheTate1998, Burger2003, Tan2007, Bonewald2008, Adachi2009}]. 

Mechanical adaptation also relies on the comparison of the current mechanical state with a reference state. The definition of this reference state remains unclear [\cite{Frost1987, Carter2001}]. Our choice is to take as mechanical reference state the initial distribution of the strain energy density in the midshaft femur. This choice introduces a memory of the stimulus ``normally" experienced in a certain region of the tissue. This memory effect leads to a position-dependent reference state which can be interpreted as taking into account different sensitivities of the mechano-sensing cells depending on where they are located [\cite{Skerry1988, Turner2002, Robling2006}].

\subsubsection*{Neutral axis and site-specific bone adaptation}
A common issue in models of mechanical adaptation is the risk to resorb too much bone in regions that are naturally unloaded. Such regions may exist when bending moment is large enough with respect to compressive or tensile forces. In the human midshaft femur, a neutral axis runs approximately along the antero-posterior axis [\cite{Lanyon1984, Cordey1999, Thomas2005, Martelli2014}]. To prevent excessive resorption in such regions, some models have considered torsional loads [\cite{Meulen1993, Carpenter2008}], average values of periodic dynamic loads under which the neutral axis moves [\cite{Meulen1993, Carter2001}], or a residual background of mechanical stimulus modelling muscle twitching and other background mechanical forces [\cite{Mittlmeier1994, Carpenter2008}]. 

Such additional features were not introduced explicitly in our model. The strength of the mechanical stimulus around the neutral axis remained weak in our simulation of mechanical disuse (Figure \ref{Model_check}(a)). This is due to the fact that stimulus sensitivity is prescribed according to the initial state. The neutral axis did not move substantially during the simulation, and so the difference in strain energy density remained small. Resorption around the neutral axis was pronounced in our simulation of osteoporosis (Figures \ref{Model_check}(b), (c) and \ref{all_plots_mecha}) due to hormone-induced remodelling imbalance. Resorption was limited by the duration of the simulated osteoporosis (40 years) and the calibration of the overall bone loss according to experimental data.

The bending moment exerted onto the femur at the midshaft creates a strong asymmetry in the local mechanical state. Over time, this asymmetry leads to a cortical wall thickness which differs between the $y$-axis (antero--posterior axis) and the $z$-axis (lateral--medial axis), as seen in Figure \ref{all_plots_mecha}(a). Asymmetries in cortical wall thickness and bone volume fraction, in osteoporotic patients, are commonly observed (Figure \ref{Model_check}(d)) [\cite{Feik2000, Thomas2005, Zebaze2010}].

\subsubsection*{Microscopic vs tissue-scale mechanical regulation}
Mechanical deformations of bone matrix can be sensed by osteocytes at the microscopic, cellular scale by deformation of the cell body, transmitted either through direct contact with the matrix, or through changes in fluid flow or hydrostatic pressure [\cite{Weinbaum1994, KnotheTate2003, Adachi2009, Adachi2009a, Bonewald2011}]. However, osteocytes are highly connected to one another and to other cells in the vascular phase by an extensive network [\cite{Marotti2000, Kerschnitzki2013, Buenzli2015}]. Whilst signal transmission mechanisms in this network remain to be determined, it is possible that the network integrates deformations of both the matrix and vascular phases before transducing them into a biochemical response, enabling a mechanical sensitivity of the network to tissue-average stresses and strains.

The uncertainty of the scale at which mechanical stimulus is sensed in bone has motivated many computational studies to estimate stress concentration effects in bone microstructures [\cite{Hipp1990, Kasiri2008, Gitman2010}]. However, few studies have explored the changes that occur during simulated bone loss when using microscopic or tissue-scale mechanical stimulus.

Our simulations of osteoporosis show that most of the difference between the mechanical stimulus at the microscopic and tissue scales occurs near the endosteum and neutral axis (Figure \ref{all_plots_mecha}(b,c)). Changes in bone volume fraction were similar in both simulations. Stress concentration effects captured in the microscopic mechanical stimulus (but not in the tissue-scale mechanical stimulus) resulted in maintaining a region of low porosity ($\fbm \simeq 0.1$) near the medullary cavity and in widening the transition between endocortical and intracortical bone volume fractions (Figure \ref{without_with_mecha}).

Osteoporotic human femur midshafts exhibit a wide range of variability, reflecting the multiple factors influencing bone loss [\cite{Feik1997, Feik2000, Thomas2005, Zebaze2010}]. The expansion of the medullary cavity and thinning of the cortical wall are commonly reported, but other changes in midshaft tissue microstructures have been studied less systematically. Depending on the subject and their specific condition, the transition between porous endocortical bone and dense intracortical bone may be sharp or wide, and highly porous microstructures near the endosteum may be found or not [\cite{Feik1997}, Figure 6]. 

Our model possesses several limitations which prevent at this stage to draw definite conclusions about the mechanical regulation of the tissue. The mechanical state is calculated only based on bone volume fraction. Other microstructural parameters such as the connectivity of the microstructure are not accounted for. Loss of connectivity is observed in osteoporotic trabecular bone [\cite{Parfitt1987, Mosekilde1990, Raisz2001}], which could lead to mechanical disuse and so increase in resorption. Periosteal apposition is often reported and believed to result from a compensation of endocortical bone loss in osteoporotic patients [\cite{Szulc2006, Russo2006, Jepsen2012}]. Our simulations assumed the periosteal surface to be fixed, which could limit the expansion rate of the medullary cavity. Finally, our simulation of osteoporosis assumed a constant level of physical activity. A reduction in physical activity with age could further limit the preservation of bone matrix by mechanical feedback.

\section*{Summary and conclusions}
In this paper a novel spatio-temporal multiscale model of bone remodelling is proposed. This model bridges organ, tissue and cellular scales. It takes into account biochemical, geometrical, and biomechanical feedbacks. The model is applied to simulate the evolution of a human femur midshaft scan under mechanical disuse and osteoporosis. It enables us to investigate how these scales and feedbacks interact during bone loss. Our numerical simulations revealed the following findings:
\begin{itemize}
\item Endocortical bone loss during both osteoporosis and mechanical disuse is driven to a large extent by site-specific turnover rates. 
\item Mechanical regulation does not influence significantly the expansion rate of the medul\-lary cavity.
\item  Mechanical regulation helps preserve cortical bone near the periosteum. It explains site-specific differences in the bone volume fraction distribution in the midshaft cross section during osteoporosis such as increased porosity near the neutral axis, and thicker cortical wall along the medial--lateral axis of the femur midshaft, due to the anisotropy of the mechanical stimulus in the presence of bending moments.
\item The inclusion of stress concentration effects in the mechanical stimulus sensed by the bone cells has a pronounced effect on porosity in the endocortical region.
\end{itemize}
Our methodology provides a framework for the future development of patient-specific models to predict loss of bone with age or deregulating circumstances.

\appendix
\section{Complements on the model description}
\subsection{Differentiation rates and signalling molecules in the cell populations model}\label{appendix_molecule_formulation}
In Section \ref{section_bone_cell_pop_methods}, we presented the simplified equations of the bone cells population model. Here are the developments of these equations.
\begin{align}
    &\docu\big(\mcsf, \rankl(\Psi,\pth)\big) = \nonumber \\
    & ~~~~~~~~~~~~~~~~~~~D_\ocu\piact\big(\tfrac{\textrm{MCSF}}{k^{\textrm{MCSF}}_{\textrm{OC}_{\textrm{u}}}}\big)\piact\big(\tfrac{\textrm{RANKL}}{k^{\textrm{RANKL}}_{\textrm{OC}_{\textrm{u}}}}\big),\notag
    \\&\docp\big(\rankl(\Psi,\pth)\big) = D_\ocp\piact\big(\tfrac{\textrm{RANKL}}{k^{\textrm{RANKL}}_{\textrm{OC}_{\textrm{p}}}}\big),\notag
    \\&\aoca(\tgfb) = A_\oca\piact\big(\tfrac{\textrm{TGF$\beta$}}{k^{\textrm{TGF$\beta$}}_{\textrm{OC}_{\textrm{a}}}}\big),\notag
    \\&\dobu(\tgfb) = D_\obu\piact\big(\tfrac{\textrm{TGF$\beta$}}{k^{\textrm{TGF$\beta$}}_{\textrm{OB}_{\textrm{u}}}}\big),\notag
    \\&\dobp(\tgfb) = D_\obp\pirep\big(\tfrac{\textrm{TGF$\beta$}}{k^{\textrm{TGF$\beta$}}_{\textrm{OB}_{\textrm{p}}}}\big).\label{governing-eqs-rates}
\end{align}
In those equations, several signalling molecules play a role: \tgfb, \rank, \rankl, \opg, \mcsf ~and \pth. The concentrations of these molecules follow the principles of mass action kinetics of receptor-ligand reactions. Due to the separation of scale between the cells differentiation and apoptosis rates and the receptor-ligand binding reactions, we solve them in a quasi-steady-state hypothesis:
\begin{align}
    \textrm{PTH}(\b r, t) = &\begin{cases}
     \textrm{P}_{\textrm{PTH}}, &\quad \text{without deregulation}
    \\\textrm{P}_{\textrm{PTH}}^{\textrm{OP}}, &\quad \text{whith simulated OP}
    \end{cases},  \label{pth}
    \\\tgfb(\b r, t) = &\frac{P^{\textrm{ext}}_{\textrm{TGF$\beta$}} + n^{\textrm{bone}}_{\textrm{TGF$\beta$}} ~\kres ~\textrm{OC}_{\textrm{a}}}{D_{\textrm{TGF$\beta$}}}
    \\\textrm{RANK}(\b r, t) = &N^{\textrm{RANK}}_{\textrm{OC}_{\textrm{p}}}\, \ocp, 
    \\\textrm{OPG}(\b r, t) = &\frac{P_{\textrm{OPG}} + \beta^{\textrm{OPG}}_{\textrm{OB}_{\textrm{a}}}\, \oba\, \pirep\!\Big(\frac{\textrm{PTH}}{k^{\textrm{PTH}}_{\textrm{OB}}}\Big)}{\beta^{\textrm{OPG}}_{\textrm{OB}_{\textrm{a}}}\,\oba\,\pirep\!\Big(\frac{\textrm{PTH}}{k^{\textrm{PTH}}_{\textrm{OB}}}\Big)/\opg_\text{sat} + D_{\textrm{OPG}}} \label{opg}
    \\\textrm{RANKL}(\b r, t) = &\frac{\beta^{\textrm{RANKL}}_{\textrm{OB}_{\textrm{p}}}\,\obp + \beta_{\textrm{RANKL}}^{\textrm{mech}}(\Psi)}{1+k^{\textrm{RANKL}}_{\textrm{RANK}}\,{\textrm{RANK}} + k^{\textrm{RANKL}}_\textrm{OPG}\,\opg} \label{RANKL}
    \\\times &\left\{\rule{0pt}{4ex}\right. \!\!D_{\textrm{RANKL}} + \frac{\beta^{\textrm{RANKL}}_{\textrm{OB}_{\textrm{p}}}\,\obp}{N^{\textrm{RANKL}}_{\textrm{OB}_{\textrm{p}}}\obp\ \piact\Big(\frac{\textrm{PTH}}{k^{\textrm{PTH}}_{\textrm{OB}}}\Big)}\!\!\left. \rule{0pt}{4ex}\right\}^{-1}.\notag
\end{align}

\subsection{Parameter values}\label{table_section}
See Table \ref{nomenclature}.
\renewcommand{\arraystretch}{1.5}
\begin{table*}
\caption{Nomenclature}
\resizebox{\textwidth}{!}{%
\begin{tabularwithnotes}{lll}
{
  \tnote[4]{Note that in comparison with \cite{Scheiner2013}, the $x$- and $z$-axes are switched.}
  \tnote[5]{Unless otherwise specified, parameter values are taken from [\cite{Buenzli2013a}]}
 }
\hline
\textbf{Symbol} & \textbf{Description} & \textbf{Value}\\
\hline
$\chi_\bv$ & Turnover rate & Extrapolated function of $f_{bm}$ from [\cite{Parfitt1983}]
\\ $\obu$ & Uncommitted osteoblasts & Given function of $f_{bm}$, determined to fulfil the steady state
\\ $\ocu$ & Uncommitted osteoclasts & Given function of $f_{bm}$, determined to fulfil the steady state 
\\ $\kform$ & Daily volume of bone matrix formed per osteoblast & 150 $\mu m^{3}$/day [\cite{Buenzli2014a}]
\\ $\kres$ & Daily volume of bone matrix resorbed per osteoclast & 9.43$\cdot 10^{3} \mu m^{3}$/day [\cite{Buenzli2014a}]
\\ $\lambda$ & Strength of the mechanical transduction in formation & 0.5 (Parametric study) 
\\ $\kappa$ & Strength of the mechanical transduction in resorption & 18 pM/day (with $\mu_\bm^\micro$), 19 pM/day (with $\mu^\tissue$) (Parametric study)
\\ $\cmicro_\bm$ & Stiffness tensor of the bone matrix phase &  \[ \left(\renewcommand{\arraystretch}{1} \begin{array}{cccccc}
28.4 & 11.0 & 10.4 & 0 & 0 & 0 \\
11.0 & 20.8 & 10.3 & 0 & 0 & 0 \\
10.4 & 10.3 & 18.5 & 0 & 0 & 0 \\
0 & 0 & 0 & 12.9 & 0 & 0 \\
0 & 0 & 0 & 0 & 11.5 & 0 \\
0 & 0 & 0 & 0 & 0 & 9.3 \end{array} \right) \textrm{GPa [\cite{Ashman1984, Fritsch2007}]\tmark[4]}\] 
\\ $\cmicro_\vas$ & Stiffness tensor of the vascular phase & \[ 2.3 \cdot \left(\renewcommand{\arraystretch}{1} \begin{array}{cccccc}
1 & 1 & 1 & 0 & 0 & 0 \\
1 & 1 & 1 & 0 & 0 & 0 \\
1 & 1 & 1 & 0 & 0 & 0 \\
0 & 0 & 0 & 0 & 0 & 0 \\
0 & 0 & 0 & 0 & 0 & 0 \\
0 & 0 & 0 & 0 & 0 & 0 \end{array} \right) \textrm{GPa [\cite{Murdock1996}]}\]  
\\ $N_{x}$ & Normal force & -700 N (see Section \ref{methods_organ_to_cellular})
\\ $M$ & Bending moment & 50 Nm (see Section \ref{methods_organ_to_cellular})
\\ & & 
\\ $D_\ocu$ & Differentiation rate of $\ocu$ into $\ocp$ & 0.42/day [\cite{Pivonka2013}]
\\ $D_\obu$ & Differentiation rate of $\obu$ into $\obp$ & 0.7/day  \tmark[5]
\\ $D_\ocp$ & Differentiation rate of $\ocp$ into $\oca$ & 2.1/day
\\ $D_\obp$ & Differentiation rate of $\obp$ into $\oba$ & 0.166/day
\\ $P_\obp$ & Proliferation term of $\obp$ &  3.5$\cdot 10^{-3}$/day
\\ $A_\oca$ & Apoptosis rate of $\oca$ & 5.65/day
\\ $A_\oba$ & Apoptosis rate of $\oba$ & 0.211/day
\\ & & 
\\ $k^{\textrm{TGF$\beta$}}_{\textrm{OC}_{\textrm{a}}}$ & Parameter for $\tgfb$ binding on $\obu$ and $\oca$ & 5.63$\cdot 10^{-4}$ pM
\\ $k^{\textrm{TGF$\beta$}}_{\textrm{OB}_{\textrm{p}}}$ & Parameter for $\tgfb$ binding on $\obp$ & 1.89$\cdot 10^{-3}$ pM
\\ $k^{\textrm{PTH}}_{\textrm{OB}}$ & Parameter for $\pth$ binding on \ob ~(activator) & 150 pM
\\ $k^{\textrm{PTH}}_{\textrm{OB}}$ & Parameter for $\pth$ binding on \ob ~(repressor) & 0.222 pM 
\\ $k^{\textrm{RANKL}}_{\textrm{OC}_{\textrm{p}}}$ & Parameter for $\rankl$ binding on $\ocp$ & 16.65 pM 
\\ $N^{\textrm{RANK}}_{\textrm{OC}_{\textrm{p}}}$ & Number of $\rank$ receptors per $\ocp$ & 1$\cdot 10^{4}$
\\ $k^{\textrm{MCSF}}_{\textrm{OC}_{\textrm{u}}}$ & Parameter for $\mcsf$ binding on $\ocu$ & 1$\cdot 10^{-3}$ pM [\cite{Pivonka2013}]
\\ $\textrm{P}_{\textrm{PTH}}$ & Systemic concentration of \pth & 2.907 pM
\\ $\textrm{P}_{\textrm{PTH}}^{\textrm{OP}}$ & $\textrm{P}_{\textrm{PTH}}$ when simulated osteoporosis & 2.954 pM
\\ $D_{\textrm{OPG}}$ & Degradation rate of \opg & 0.35/day
\\ $\beta^{\textrm{OPG}}_{\textrm{OB}_{\textrm{a}}}$ & Production rate of \opg ~per $\oba$ & 1.63$\cdot 10^{8}$/day
\\ $\opg_{\textrm{sat}}$ & Saturation of \opg & 2$\cdot 10^{8}$ pM 
\\ $D_{\textrm{RANKL}}$ & Degradation rate of \rankl & 10/day
\\ $\beta^{\textrm{RANKL}}_{\textrm{OB}_{\textrm{p}}}$ & Production rate of \rankl ~per $\obp$ & 1.68$\cdot 10^{5}$/day 
\\ $k^{\textrm{RANKL}}_{\textrm{OPG}}$ & Parameter for \rankl ~binding on \opg & 1$\cdot 10^{-3}$/pM 
\\ $k^{\textrm{RANKL}}_{\textrm{RANK}}$ & Parameter for \rankl ~binding on \rank & 0.034/pM
\\ $D_{\textrm{TGF$\beta$}}$ & Degradation rate of \tgfb & 2/day
\\ $n^{\textrm{bone}}_{\textrm{TGF$\beta$}}$ & Density of \tgfb ~stored in the bone matrix & 1$\cdot 10^{-2}$ pM  \\
\hline
\end{tabularwithnotes}}
\label{nomenclature}
\end{table*}

\subsection{Recalibration of the model}\label{appendix_calib}
Since $\ocu$ and $\obu$ vary with $\fbm$ so as to retrieve experimentally valid turnover rates, some other parameters required modification compared with previous versions of the cell population model in which $\ocu$ and $\obu$ were constant and uncalibrated~[\cite{Buenzli2013a, Pivonka2013}].

By comparing the cell densities between this model and the previously published one [\cite{Buenzli2013a}], we can determine scaling coefficients which allows a systematic calibration of $\piact\big(\tfrac{\textrm{TGF$\beta$}}{k^{\textrm{TGF$\beta$}}_{\textrm{OC}_{\textrm{a}}}}\big)$  and $\piact\big(\tfrac{\textrm{RANKL}}{k^{\textrm{RANKL}}_{\textrm{OC}}}\big)$. Indeed, these functions depend on the active and precursor cell densities. In the original models, the constants in these functions were calibrated such as to obtain a strong biochemical feedback response. Maintaining this strong biochemical response is the aim of this re-calibration.

The calibration is realised at \fbm = 0.90. Both the turnover rate value and the values of $\kres$ and $\kform$, have been changed according to the literature. Hence, by isolating $\textrm{OB}_{\textrm{a}}$ and $\textrm{OC}_{\textrm{a}}$ in the two new constraints of the steady state, the active osteoblast and active osteoclast read: 
\begin{equation}
\textrm{OC}_{\textrm{a}}^{\textrm{new}} = \frac{\chi_\bv^{\textrm{new}}}{k_{\textrm{res}}^{\textrm{new}}} = \beta \cdot \textrm{OC}_{\textrm{a}}
\end{equation} 
\begin{equation}
\textrm{OB}_{\textrm{a}}^{\textrm{new}} = \frac{k_{\textrm{res}}^{\textrm{new}} \cdot \textrm{OC}_{\textrm{a}}^{\textrm{new}}}{k_{\textrm{form}}^{\textrm{new}}} = \gamma \cdot \textrm{OB}_{\textrm{a}}
\end{equation} 
if $\delta$ is the coefficient of proportionality between the new bone turnover rate and the previous one; $\beta = \delta \cdot \kres / k_{\textrm{res}}^{\textrm{new}}$ and $\gamma = \delta \cdot \kform / k_{\textrm{form}}^{\textrm{new}}$. These coefficients are introduced in the determination of $\textrm{TGF$\beta$}$ and $\textrm{OPG}$. Previously, $\textrm{TGF$\beta$}$ was [\cite{Buenzli2013a}]:
\begin{equation}
\textrm{TGF$\beta$} = \frac{P^{\textrm{ext}}_{\textrm{TGF$\beta$}} + n^{\textrm{bone}}_{\textrm{TGF$\beta$}} ~\kres ~\textrm{OC}_{\textrm{a}}}{D_{\textrm{TGF$\beta$}}} 
\end{equation} 
The new one becomes:
\begin{equation}
\textrm{TGF$\beta$}^{\textrm{new}} = \frac{P^{\textrm{ext}}_{\textrm{TGF$\beta$}} + n^{\textrm{bone}}_{\textrm{TGF$\beta$}} ~k_{\textrm{res}}^{\textrm{new}} ~\textrm{OC}_{\textrm{a}}^{\textrm{new}} \cdot \delta^{-1}}{D_{\textrm{TGF$\beta$}}}
\end{equation} 
The same manipulation is realised on the determination of \opg. The factor $\beta^{\textrm{OPG}}_{\textrm{OB}_{\textrm{a}}}\, \textrm{OB}_{\textrm{a}}$ in Eq. \eqref{opg}, becomes $\beta^{\textrm{OPG}}_{\textrm{OB}_{\textrm{a}}}\, \textrm{OB}_{\textrm{a}}^{\textrm{new}} \gamma^{-1}$.

\section{Update frequency of mechanical state in the numerical algorithm}\label{appendix_delta_t}
In our model, to simulate osteoporosis and the change of porosity with time, we need to solve the temporal equations of the bone cell populations model, Eqs (\ref{dOCpdt})--(\ref{dOBadt}) and Eq. (\ref{fbm}). Those equations via the mechanical feedback are correlated to the spatial Equations (\ref{Bernoulli}) and (\ref{deformation}). Knowing the porosity distribution is required to determine the stress and strain distributions. Hence we have a semi-coupled algorithm (Figure \ref{model}).

However, due to the separation of time scale we can decompose the problem into two parts. Indeed, it takes more time for the microstructure to change significantly enough to influence the bone cell populations model. Therefore, we solve the bone cell populations model for a duration $\Delta$t, assuming the mechanical feedback to be constant in this time interval. Then, we recalculate the stress and strain distributions based on the new porosity distribution, and this becomes the new mechanical feedback.  

\begin{figure}[h]
\begin{center} 
\includegraphics[trim = 0.5cm 16.3cm 5cm 1.1cm, clip=true, width = 0.4\textwidth]{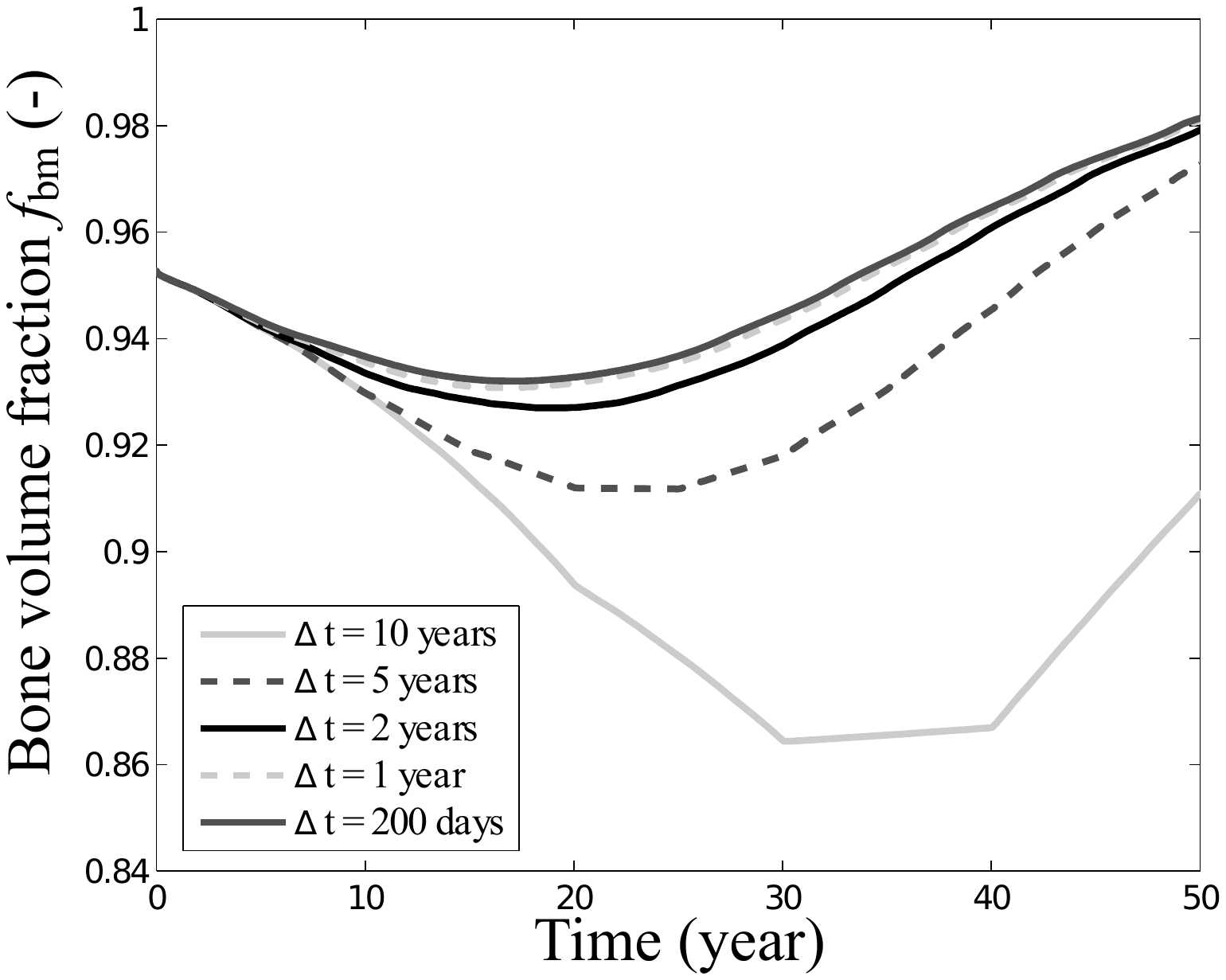}
\caption{The evolution of the bone matrix volume fraction for different time steps. Note that this \rve ~is in the intracortical region which undergoes first resorption then formation due to the redistribution of the mechanical loads.}
\label{Delta_Time}
\end{center}
\end{figure}

A sensitivity analysis of the solution in the time step $\Delta$t of evolution of cell densities and bone matrix volume fraction is required. For very small time steps ($\Delta$t $\leq$ 1 day) one would expect the algorithm to converge to the exact solution. On the other hand for very large time steps ($\Delta$t $\geq$ 5 years) a large deviation from the exact solution is expected. Figure \ref{Delta_Time} shows the evolution of the bone matrix volume fraction for one selected RVE ($y$ = 0, $z$ = -10 mm) in the cross section. These simulations show that time steps of $\Delta$t = 250 days, 1 year and 2 years lead to very similar evolution of the bone matrix volume fraction. On the other hand, $\Delta$t = 5 years and 10 years lead to strong deviations from the smaller time increments. For all the simulations of 40 years of osteoporosis, we used a time step of 2 years.

\section{Generalised Beam theory for inhomogeneous materials}\label{appendix:beam}

In the following, we represent the governing equations using a Cartesian ($x$, $y$, $z$) coordinate system. The $x$-axis represents the beam axis and coincides with the direction of the vascular pores (i.e., Haversian systems). The $y$ and $z$ coordinates describe a material point in the cross section (Figure \ref{multiscale}(c)). The origin of the system is known as the \textit{Normal force Center}: NC. Since our cross sections are inhomogeneous all the quantities, including the stiffness, are functions of $y$ and $z$. 

First, based on the constitutive relation: Hooke's law, we determine the strain and stress relation:
\begin{align}\label{sigma}
\b \sigma^\tissue(y, z, t) = \Ctissue(y, z, t) : \b \varepsilon^\tissue(y, z, t)
\end{align}
where $\b \sigma^\tissue(y, z, t)$ and $\b \varepsilon^\tissue(y, z, t)$ are the ``tissue" stress and strain and $\Ctissue(y, z, t)$ the tissue stiffness matrix. The stiffness matrix is determined at the tissue scale, the explanation is presented in Section \ref{methods_organ_to_cellular}. 

Based on the Bernoulli hypothesis, the strain distribution appears to be a plane and remains plane even after deformation. This is why we can decompose the strain by introducing three constants: $\varepsilon_1$, $\kappa_3$ and $\kappa_2$. 
\begin{align}\label{beam_theory_strain}
    \varepsilon^\tissue_{xx}(y,z,t) = \varepsilon_1(t) - \kappa_3(t) y + \kappa_2(t) z.
\end{align}
By introducing this relation into Hooke's law, we obtain:
\begin{equation} \label{beam_theory_stress}
\sigma^\tissue_{xx}(y, z, t) = \Ctissue_{xx}(y, z, t)~(\varepsilon_1(t) - \kappa_3(t) y + \kappa_2(t) z)
\end{equation}
Because we assume the shear force to be null, the stress tensor is reduced to one component: $\sigma^\tissue_{xx}(y, z, t)$. And with Bernoulli hypothesis the strain tensor contains only one component. Hence, the stiffness matrix can be replaced by the component $\Ctissue_{xx}(y, z, t)$. 

Here we can see that if we determine the strain constants, we would know the stress distribution. The mechanical loadings, the inputs of this model, allow us to determine the strain. Indeed the cross section is supposed to be under a normal force: $\b N$ and a bending moment $\b M$ here divided in two bending moments: $M_y$ and $M_z$, such as $M ~\hat{\b m} = M_y ~\hat{\b y} + M_z ~\hat{\b z}$. By definition of the stress we have the relations: 
\begin{align}
N&= \int \sigma^\tissue_{xx}(y, z, t) dA \\
M_y &= \int z \cdot \sigma^\tissue_{xx}(y, z, t) dA \\
M_z &= - \int y \cdot \sigma^\tissue_{xx}(y, z, t) dA
\end{align}
By introducing the static moments of first and second order: $EA$, $ES_{y}$, $ES_{z}$, $EI_{yy}$, $EI_{zz}$, $EI_{yz}$, the equations become the following constitutive relation: 
\begin{equation}
\begin{bmatrix}
   N \\
   M_y \\
   M_z
\end{bmatrix}
 = 
\begin{bmatrix}
   EA & ES_{y} & ES_{z} \\
   ES_{y} & EI_{yy} & EI_{yz} \\
   ES_{z} & EI_{yz} & EI_{zz}
\end{bmatrix}
\begin{bmatrix}
   \varepsilon_1 \\
   \kappa_2 \\
   \kappa_3
\end{bmatrix}
\label{deformation}
\end{equation}
where:
\begin{alignat*}{4}
&EA &&= \int \Ctissue_{xx}(y, z, t) dA\\
&ES_{y} &&= \int \Ctissue_{xx}(y, z, t) \cdot y dA\\
&EI_{yy} &&= \int \Ctissue_{xx}(y, z, t) \cdot y^{2} dA\\
&ES_{z} &&= \int \Ctissue_{xx}(y, z, t) \cdot z dA\\
&EI_{zy} &&= \int \Ctissue_{xx}(y, z, t) \cdot yz dA = EI_{yz}\\
&EI_{zz} &&= \int \Ctissue_{xx}(y, z, t) \cdot z^{2} dA
\end{alignat*}

If we chose the origin of the coordinate system at the normal center (NC) of the cross section, the coupling terms between extension and bending vanish since they become null by definition of the NC:
\begin{equation}
ES_{y} = \int \Ctissue_{xx}(y, z, t) \cdot y dA = 0
\end{equation}
\begin{equation}
ES_{z} = \int \Ctissue_{xx}(y, z, t) \cdot z dA = 0
\end{equation}
The constitutive relation can be simplified as:
\begin{align}
\begin{bmatrix}
   N \\
   M_{y} \\
   M_{z}
\end{bmatrix}
 = 
\begin{bmatrix}
   EA & 0 & 0 \\
   0 & EI_{yy} & EI_{yz} \\
   0 & EI_{yz} & EI_{zz}
\end{bmatrix}
\begin{bmatrix}
   \varepsilon_1 \\
   \kappa_2 \\
   \kappa_3
\end{bmatrix}
\label{deformation}
\end{align}

\subsection*{Determination of the Normal Force Center – NC}
The special location of the origin of the coordinate system for which the coupling terms ($ES_{y}$ and $ES_{z}$) between extension and bending become zero is by definition called the normal force center NC. The result of this definition is that an axial force $\b N$ which acts at the NC only causes straining and no bending. The coupling terms are also referred to as weighted static moments or weighted first order moments. To find the position of the coordinate system for which the coupling terms become zero requires a tool.

Assume a temporary coordinate system: $\bar{y} - \bar{z}$ from which the porosity distribution is known. The shift in origin of this coordinate system with respect to the $y$ - $z$ coordinate system through the unknown NC is denoted with $\bar{y}_{NC}$ and $\bar{z}_{NC}$. The temporary coordinate system can be expressed in terms of the $y$ - $z$ coordinate system as:
$$\bar{y} = y + \bar{y}_{NC} ~~~~~~~~ \bar{z} = z + \bar{z}_{NC}$$
Hence:
\begin{align}
ES_{\bar{y}} &= \int \Ctissue_{xx}(y, z, t) \cdot \bar{y} dA \\
&= \int \Ctissue_{xx}(y, z, t) \cdot y dA + \bar{y}_{NC} \int \Ctissue_{xx}(y, z, t) dA \nonumber \\
&= ES_{y} + EA \cdot \bar{y}_{NC} \nonumber\\\nonumber
\\ES_{\bar{z}} &= \int \Ctissue_{xx}(y, z, t) \cdot \bar{z} dA \\
&= \int \Ctissue_{xx}(y, z, t) \cdot z dA + \bar{z}_{NC} \int \Ctissue_{xx}(y, z, t) dA \nonumber \\
&= ES_{z} + EA \cdot \bar{z}_{NC} \nonumber
\end{align}
By definition $ES_{y}$ and $ES_{z}$ with respect to the $y$ - $z$ coordinate system are zero. From which the unknown position of the NC with respect to the known position of the $\bar{y} - \bar{z}$ coordinate system can be found:
\begin{equation}
\bar{y}_{NC} = \frac{ES_{\bar{y}}}{EA}
\end{equation}
\begin{equation}
\bar{z}_{NC} = \frac{ES_{\bar{z}}}{EA}
\end{equation}

To conclude, here is the step-by-step methodology we are using to find the stress and strain distribution in the cross section:
\begin{enumerate}
\item Localise the normal center (NC) by computing the integrations: $EA$, $ES_{y}$, $ES_{z}$.
\item Compute the integrations: $EI_{yy}$, $EI_{zz}$ and $EI_{yz}$.
\item Determine the cross-sectional forces: $\b N$, $M_{y}$ and $M_{z}$.
\item Calculate the cross-sectional deformations: $\varepsilon_1$, $\kappa_2$ and $\kappa_3$ based on Eqn. (\ref{deformation}).
\item Find the strain distribution based on the kinematic relation, Eqn. (\ref{beam_theory_strain}). Here it is important to remember to use the coordinate centred in NC.
\item Find the stress distribution based on Hookes' law, Eqn. (\ref{sigma}).
\end{enumerate}

The initial cross section is extracted from a microradiograph, as it is explained in Section \ref{microCT_section}; and the mechanical loading is not symmetrical. Hence the position of the NC is changing. This is why we need to localise it after each step.

\paragraph{}
\textbf{Conflict of Interest} The authors declare that they have no conflict of interest.

\begin{acknowledgements}
We thank C. David L. Thomas and Prof. John G. Clement for providing the microradiographs of the femur cross sections. PRB is the recipient of an Australian Research Council Discovery Early Career Research Award (DE130101191).
\end{acknowledgements}

\bibliographystyle{spbasic}      
\bibliography{My-Collection_v1}   

\end{document}